\pdfoutput=1
\documentclass[12pt]{article}
\usepackage[margin=1in]{geometry}
\usepackage{amsmath}
\usepackage{titling}
\usepackage{blindtext}
\usepackage{pgfplots}
\usepackage{natbib}
\usepackage{comment}
\usepackage[colorlinks=true,linkcolor=blue,allcolors=blue]{hyperref}
\usepackage{url}
\usepackage{setspace}
\usepackage{authblk}
\pgfplotsset{compat=1.15}

\providecommand{\keywords}[1]
{
  \small	
  \textit{Keywords:} #1
}

\begin{document}

\title{Coronal Loop Heating by Nearly Incompressible Magnetohydrodynamic and Reduced Magnetohydrodynamic Turbulence Models}


\author[1]{M. S. Yalim}
\affil[1]{Center for Space Plasma and Aeronomic Research, The University of Alabama in Huntsville, Huntsville, AL 35805, USA}

\author[1,2]{G. P. Zank}
\affil[2]{Department of Space Science, The University of Alabama in Huntsville, Huntsville, AL 35805, USA}

\author[3]{M. Asgari-Targhi}
\affil[3]{Harvard-Smithsonian Center for Astrophysics, Cambridge, MA 02138, USA}

\setcounter{Maxaffil}{0}
\renewcommand\Affilfont{\itshape\small}
\date{}    
\begin{titlingpage}
    \maketitle
\begin{abstract}
The transport of waves and turbulence beyond the photosphere is central to the coronal heating problem. Turbulence in the quiet solar corona has been modeled on the basis of the nearly incompressible magnetohydrodynamic (NI MHD) theory to describe the transport of low-frequency turbulence in open magnetic field regions. It describes the evolution of the coupled majority quasi-2D and minority slab component, driven by the magnetic carpet and advected by a subsonic, sub-Alfv\'enic flow from the lower corona. In this paper, we couple the NI MHD turbulence transport model with an MHD model of the solar corona to study the heating problem in a coronal loop. In a realistic benchmark coronal loop problem, we find that a loop can be heated to $\sim$1.5 million K by transport and dissipation of MHD turbulence described by the NI MHD model. We also find that the majority 2D component is as important as the minority slab component in the heating of the coronal loop. We compare our coupled MHD/NI MHD model results with a reduced MHD (RMHD) model. An important distinction between these models is that RMHD solves for small-scale velocity and magnetic field fluctuations and obtains the actual viscous/resistive dissipation associated with their evolution whereas NI MHD evolves scalar moments of the fluctuating velocity and magnetic fields and approximates dissipation using an MHD turbulence phenomenology. Despite the basic differences between the models, their simulation results match remarkably well, yielding almost identical heating rates inside the corona.
\end{abstract}
\end{titlingpage}



\section{Introduction} \label{sec:intro}

The plasma temperature from the photosphere to corona increases from $\sim$5,000 K to $\sim$1 million K over a distance of only $\sim$10,000 km from the chromosphere and the transition region to the corona. Understanding the mechanism underlying coronal heating is a fundamental problem in the solar physics community. The transport of waves and turbulence beyond the photosphere is central to the coronal heating problem~\citep{Matthaeus99,Oughton2001,CvB10,van Ballegooijen11,Cranmer15,vBTA16,vBTA17,Zank18,Zank21}.

In a coronal loop, Alfv\'en waves are generated along the loop by dynamic transverse twisting and braiding motions in its footpoints on the photosphere where magnetic flux tubes are distorted by convective flows in intergranular lanes~\citep{van Ballegooijen11}. These Alfv\'en waves then propagate outward along the magnetic field lines and dissipate their energy in the chromosphere and corona. During this process, due to the gradients in the outward-propagating Alfv\'en wave velocities, inward-propagating modes are generated resulting in complex counter-propagating interactions between these Alfv\'en waves. A key insight
introduced by~\citet{Matthaeus99} is that the outward-propagating and reflected inward-propagating Alfv\'en waves couple non-linearly through the production of 2D fluctuations~\citep{Shebalin83}, i.e., zero-frequency non-propagating fluctuations that undergo a rapid 2D ($\textbf{$\emph{k}_{\perp}$}$, perpendicular to the mean magnetic field $\textbf{$\emph{B}_{0}$}$) turbulent cascade (successive reconnection of quasi-2D or poloidal magnetic flux structures) that transfers energy to progressively smaller perpendicular scales until it dissipates at presumably ion inertial/gyrofrequency scales~\citep{Matthaeus99,Oughton2001,CvB10,Cranmer15,Zank18}. 

Many previous studies involving numerical simulations describe the loop heating mechanism by turbulent relaxation of braided/tangled magnetic field structures whether initially present or built up in the course of the simulation~\citep[e.g.][]{Dahlburg12,RP13,PH15,Wilmot-Smith15,Pontin17, Pontin20} without actually solving the turbulence transport equations. In particular,~\citet{RP13} investigate formation of current sheets in tangled magnetic field structures following the coronal heating mechanism due to nanoflares~\citep{Parker88}. The \cite{RP13} simulation, however, offers a quite different perspective on the heating problem in coronal loops compared to the counter-propagating Alfv\'en wave picture described above. Instead, Rappazzo \& Parker use randomized 2D magnetic
potential to initialize the simulation that results in (their Figure 5) 2D
islands, interspersed by rapidly developing current sheets.  Not surprisingly, in the presence of a strong guide magnetic field, the fluctuating fields are dominated by 2D structures rather than counter-propagating Alfv\'en waves. Such a mechanism for loop heating closely resembles the model introduced to heat open coronal holes by \cite{CvB10,Zank18,Zank21}.

In this paper, we describe the heating mechanism in coronal loops by the nearly incompressible magnetohydrodynamic (NI MHD) turbulence transport model~\citep{Zank17}. In the NI MHD turbulence transport model, we do not explicitly introduce any transverse small scale fields or braiding of magnetic field lines as they are already accounted for by the turbulence transport equations. The magnetic
field in the immediate vicinity of the photosphere has been called the ``magnetic carpet''~\citep{TS98}. In the low plasma beta environment, transverse photospheric convective fluid motions drive predominantly 2D (non-propagating) turbulence in the mixed-polarity magnetic carpet, together with a minority slab (Alfv\'enic) component~\citep{Zank18} along the strong, uniform axial guide field inside the loop. The NI MHD model has been used in developing a turbulence-driven solar wind model for a fast solar wind flow in a coronal hole~\citep{Adhikari20} and a solar wind model that includes electron pressure and heat flux~\citep{Adhikari21}. In this paper, we focus on the coronal loop heating problem by solving the NI MHD turbulence transport model and MHD coronal model~\citep{Yalim17,Singh18} equations simultaneously in a time-dependent fashion. The MHD coronal model is utilized to solve for the background plasma in the loop. These two systems of equations are coupled via the turbulent coronal heating term in the MHD energy equation.

We compare our coupled MHD/NI MHD model results with model results from the reduced MHD (RMHD) approximation~\citep{van Ballegooijen11,AvB12}. 

The RMHD equations for a uniform background field were first derived by \citet[]{Kadomtsev1974, Strauss1976}, and studied by \citet{Montgomery1982} and \citet{Hazeltine1983} among others. \citet[]{Zank1992} extensively studied the relationships between compressible MHD, incompressible MHD, and RMHD. In the RMHD approximation, the magnetic and velocity fluctuations are assumed to be small compared to the background field and Alfv\'en speed, respectively.

The RMHD (or \textit{Alfv\'en wave turbulence}) model describes the generation, propagation and dissipation of Alfv\'en waves in a coronal loop represented by a thin flux tube surrounding the axial guide magnetic fieldline. To model the coronal loop plasma, the RMHD approximation retains only the long wavelength Alfv\'en wave modes, filtering out all fast/slow modes and the high-freq/short wavelength Alfv\'en waves. Furthermore, the magnetic and velocity fluctuations are simulated but their effects on temperature and density are ignored. Besides coronal loop heating, RMHD models have been used to model the heating of open field coronal regions e.g., \cite{Oughton2001,vBTA16,vBTA17,Asgari2021}.

Section~\ref{goveq} presents the two systems of governing equations that are coupled, namely the NI MHD turbulence transport equations to compute the coronal heating and the ideal MHD equations to calculate the background coronal plasma in the loop. Moreover, an overview of the RMHD model is also presented in this section. Section~\ref{res} presents and discusses the results obtained by the NI MHD turbulence transport model and the RMHD model. In particular, we consider a realistic benchmark coronal loop heating problem where the loop is heated to $\sim$1.5 million K from an initial uniform temperature of $8.25\times10^5$ K. Finally, section~\ref{conc} presents our conclusions.   

\section{Governing Equations} 
\label{goveq}

In this section, we first present the systems of governing equations that we solved simultaneously in a time-dependent fashion: The NI MHD turbulence transport equations and the ideal MHD equations for the MHD coronal model. We also give an overview of the RMHD model and its equations.

\subsection{NI MHD Turbulence Transport Model}
\label{NIMHD}

The system of NI MHD turbulence transport equations consists of 12 equations: 7 to describe the majority quasi-2D turbulence and the remaining 5 to describe the minority slab component. The transport variables corresponding to 2D turbulence and slab turbulence are indicated by the superscripts $\infty$ and $*$, respectively. In addition, the transport variables corresponding to forward (outward) propagating and backward (inward) propagating modes are labeled by the superscripts $+$ and $-$, respectively or sometimes by the superscripts $\pm$ or $\mp$ as a compact notation to write the transport variables with the superscripts $+$ and $-$ under a single term.

We write the 3D NI MHD model equations in differential form as a system of advection equations as follows:
\begin{equation}
\label{eq1}
\frac{\partial\textbf{\emph{U}}}{\partial t} +
\mathbf{\nabla}\cdot\textbf{\emph{F}} =
\textbf{\emph{RHS}},
\end{equation}
where $\textbf{\emph{U}}$ is the vector of turbulence transport variables which are the solution variables, $\textbf{\emph{F}}$ is the flux vector, and $\textbf{\emph{RHS}}$ is the vector of source terms which is located on the right-hand-side (RHS) of Eq.~\ref{eq1}.

$\textbf{\emph{U}}$ is given as follows:
\begin{equation}
\label{eq2}
\textbf{\emph{U}} =
\left(
\begin{array}{c c c c c c c c c}
\big<z^{\infty\pm2}\big> &
E_D^\infty &
L_\infty^\pm &
L_D^\infty &
\big<\rho^{\infty2}\big> &
\big<z^{*\pm2}\big> &
E_D^* &
L_* &
L_D^*
\end{array}
\right)^{T},
\end{equation}
where $\big< z^{\infty\pm2} \big>$, which includes $\big<z^{\infty+2}\big>$ and $\big<z^{\infty-2}\big>$, and $\big< z^{*\pm2} \big>$, which includes $\big< z^{*+2} \big>$ and $\big< z^{*-2} \big>$, are the ensemble-averaged quasi-2D and slab Els\"asser variables for backward/forward propagating modes, $E_D^\infty$ and $E_D^*$ are 2D and slab residual energy components, $L_\infty^\pm$, which includes $L_\infty^{+}$ and $L_\infty^{-}$, and $L_*$ are 2D and slab energy-weighted correlation lengths corresponding to backward/forward propagating modes, $L_D^\infty$ and $L_D^*$ are 2D and slab energy-weighted correlation lengths corresponding to residual energy, respectively, and $\big< \rho^{\infty2} \big>$ is the variance of the advected density (entropic) fluctuations. We assume $L^{+}_{*}$ = $L^{-}_{*}$ = $L_*$~\citep{Dosch13} to reduce the complexity of the transport equations for slab energy-weighted correlation lengths corresponding to backward/forward propagating modes.

We write $\textbf{\emph{F}}$ as follows:
\begin{equation}
\label{eq3}
\textbf{\emph{F}} =
\left(
\begin{array}{c c c c c c c c c}
\textbf{\emph{v}}\big<z^{\infty\pm2}\big> &
\textbf{\emph{v}}E_D^\infty &
\textbf{\emph{v}}L_\infty^\pm &
\textbf{\emph{v}}L_D^\infty &
\textbf{\emph{v}}\big<\rho^{\infty2}\big> &
\big(\textbf{\emph{v}}\mp\textbf{$\emph{v}_{A}$}\big)\big<z^{*\pm2}\big> &
\textbf{\emph{v}}E_D^* &
\textbf{\emph{v}}L_* &
\textbf{\emph{v}}L_D^*
\end{array}
\right)^{T},
\end{equation}
where $\textbf{\emph{v}}$ and $\textbf{$\emph{v}_{A}$}=\frac{\textbf{\emph{B}}}{\sqrt{4\pi\rho}}$ are the bulk (i.e., background) plasma and Alfv\'en wave velocities with $\rho$ and $\textbf{\emph{B}}$ as the bulk plasma density and magnetic field, respectively. We would like to note that $\big(\textbf{\emph{v}}\mp\textbf{$\emph{v}_{A}$}\big)\big<z^{*\pm2}\big>$ includes $\big(\textbf{\emph{v}}-\textbf{$\emph{v}_{A}$}\big)\big<z^{*+2}\big>$ and $\big(\textbf{\emph{v}}+\textbf{$\emph{v}_{A}$}\big)\big<z^{*-2}\big>$.

Finally, $\textbf{\emph{RHS}}$ can be written in the following form:
\begin{equation}
\label{eq4}
\textbf{\emph{RHS}} =
\left(
\begin{array}{c c c c c c c c c}
\big(RHS\big)_{\big<z^{\infty\pm2}\big>} &
\big(RHS\big)_{E_D^\infty} &
\big(RHS\big)_{L_\infty^\pm} &
\big(RHS\big)_{L_D^\infty} &
\big(RHS\big)_{\big<\rho^{\infty2}\big>} & \\
\big(RHS\big)_{\big<z^{*\pm2}\big>} &
\big(RHS\big)_{E_D^*} &
\big(RHS\big)_{L_*} &
\big(RHS\big)_{L_D^*}
\end{array}
\right)^{T},
\end{equation}
with
\begin{equation}
\label{eq5}
\begin{aligned}
\big(RHS\big)_{\big<z^{\infty\pm2}\big>} = &\frac{1}{2}\big<z^{\infty\pm2}\big>\mathbf{\nabla}\cdot\textbf{\emph{v}}-\Big(2a-\frac{1}{2}\Big)E_D^\infty\mathbf{\nabla}\cdot\textbf{\emph{v}}+\frac{1}{2}\big(\big<z^{\infty\pm2}\big>-E_D^\infty\big)\big<z^{\infty\pm2}\big>^\frac{1}{2}\frac{1}{\rho}\textbf{\emph{\^n}}\cdot\mathbf{\nabla}\rho \\
&-2\frac{\big<z^{\infty\pm2}\big>^2\big<z^{\infty\mp2}\big>^\frac{1}{2}}{L_\infty^\pm},
\end{aligned}
\end{equation}
where $\rho$ is the plasma density, $\textbf{\emph{\^n}}$ is an orthonormal vector orthogonal to the local large-scale mean magnetic field $\textbf{$\emph{B}_{0}$}$, and $a$ denotes a structural similarity parameter associated specifically with relating the cross-correlations of the velocity fluctuations to the 1-point velocity correlation~\citep{Zank12}, which we take as $a=1/2$;
\begin{equation}
\label{eq6}
\begin{aligned}
\big(RHS\big)_{E_D^\infty} = 
&\frac{1}{2}E_D^\infty\mathbf{\nabla}\cdot\textbf{\emph{v}}-\Big(2a-\frac{1}{2}\Big)E_T^\infty\mathbf{\nabla}\cdot\textbf{\emph{v}}+\frac{1}{4}\big(E_D^\infty-\big<z^{\infty\pm2}\big>^\frac{1}{2}\big<z^{\infty\mp2}\big>^\frac{1}{2}\big)\big(\big<z^{\infty+2}\big>^\frac{1}{2}+\big<z^{\infty-2}\big>^\frac{1}{2}\big)\frac{1}{\rho}\textbf{\emph{\^n}}\cdot\mathbf{\nabla}\rho \\
&-E_D^\infty\Big(\frac{\big<z^{\infty+2}\big>\big<z^{\infty-2}\big>^\frac{1}{2}}{L_\infty^+}+\frac{\big<z^{\infty-2}\big>\big<z^{\infty+2}\big>^\frac{1}{2}}{L_\infty^-}\Big),
\end{aligned}
\end{equation}
where $E_T^\infty=\big(\big<z^{\infty+2}\big>+\big<z^{\infty-2}\big>\big)/2$ is the total energy in 2D fluctuations;
\begin{equation}
\label{eq7}
\big(RHS\big)_{L_\infty^\pm} = 
\frac{1}{2}L_\infty^\pm\mathbf{\nabla}\cdot\textbf{\emph{v}}-\Big(a-\frac{1}{4}\Big)L_D^\infty\mathbf{\nabla}\cdot\textbf{\emph{v}}-\frac{1}{4}\big<z^{\infty\pm2}\big>^\frac{1}{2}\big(L_D^\infty-2L_\infty^\pm\big)\frac{1}{\rho}\textbf{\emph{\^n}}\cdot\mathbf{\nabla}\rho;     
\end{equation}
\begin{equation}
\label{eq8}
\begin{aligned}
\big(RHS\big)_{L_D^\infty} =
&\frac{1}{2}L_D^\infty\mathbf{\nabla}\cdot\textbf{\emph{v}}-\Big(2a-\frac{1}{2}\Big)\big(L_\infty^++L_\infty^-\big)\mathbf{\nabla}\cdot\textbf{\emph{v}}+\frac{1}{4}\Big(L_D^\infty\big(\big<z^{\infty+2}\big>^\frac{1}{2}+\big<z^{\infty-2}\big>^\frac{1}{2}\big)-2L_\infty^+\big<z^{\infty-2}\big>^\frac{1}{2} \\
&-2L_\infty^-\big<z^{\infty+2}\big>^\frac{1}{2}\Big)\frac{1}{\rho}\textbf{\emph{\^n}}\cdot\mathbf{\nabla}\rho;
\end{aligned}
\end{equation}
\begin{equation}
\label{eq9}
\big(RHS\big)_{\big<\rho^{\infty2}\big>} = 
-\big<\rho^{\infty2}\big>\mathbf{\nabla}\cdot\textbf{\emph{v}}+2\big<\rho^{\infty2}\big>\big<u^{\infty2}\big>^\frac{1}{2}\frac{1}{\rho}\textbf{\emph{\^n}}\cdot\mathbf{\nabla}\rho-\frac{\big<u^{\infty2}\big>^\frac{1}{2}\big<\rho^{\infty2}\big>}{l_{u}^\infty},
\end{equation}
where $\big<u^{\infty2}\big>=\big(E_T^\infty+E_D^\infty\big)/2$ is the kinetic energy density, and $l_{u}^\infty$ is the corresponding correlation length of the 2D velocity fluctuations given by~\citep{Zank17}
\begin{equation}
\label{eq10}
l_{u}^\infty = \frac{\big(E_T^\infty+E_C^\infty\big)\lambda_\perp^++\big(E_T^\infty-E_C^\infty\big)\lambda_\perp^-+E_D^\infty\lambda_D^\infty}{2\big(E_T^\infty+E_D^\infty\big)}
= \frac{L_\infty^++L_\infty^-+L_{D}^\infty}{2\big(E_T^\infty+E_D^\infty\big)},
\end{equation}
with $E_C^\infty=\big(\big<z^{\infty+2}\big>-\big<z^{\infty-2}\big>\big)/2$ as the 2D cross-helicity, and $\lambda_\perp^\pm=L_\infty^\pm/\big<z^{\infty\pm2}\big>$ and $\lambda_D^\infty=L_D^\infty/E_D^\infty$ as the respective correlation lengths for the 2D forward and backward energy densities for the Els\"asser variables $\big<z^{\infty\pm2}\big>$ and the 2D residual energy $E_D^\infty$; 
\begin{equation}
\label{eq11}
\begin{aligned}
\big(RHS\big)_{\big<z^{*\pm2}\big>} = 
&\frac{1}{2}\big<z^{*\pm2}\big>\mathbf{\nabla}\cdot\textbf{\emph{v}}\mp\big<z^{*\pm2}\big>\mathbf{\nabla}\cdot\textbf{$\emph{v}_{A}$}-\Big(2b-\frac{1}{2}\Big)\mathbf{\nabla}\cdot\textbf{\emph{v}}{E_D^*} + 2b{E_D^*}S_{i}S_{j}\frac{\partial v_{i}}{\partial x_{j}} \\
&\mp b{E_D^*}\Big(2\mathbf{\nabla}\cdot\textbf{$\emph{v}_{A}$}-2S_{i}S_{j}\frac{\partial v_{A_i}}{\partial x_{j}}+\frac{1}{\rho}\textbf{$\emph{v}_{A}$}\cdot\mathbf{\nabla}\rho-S_{i}S_{j}v_{A_i}\frac{1}{\rho}\frac{\partial\rho}{\partial x_{j}}\Big) \\
&\pm\frac{1}{2}\big(\big<z^{*\pm2}\big>-{E_D^*}\big)\Big[\frac{1}{\rho}\textbf{$\emph{v}_{A}$}\cdot\mathbf{\nabla}\rho\pm\big<z^{\infty\pm2}\big>^\frac{1}{2}\frac{1}{\rho}\textbf{\emph{\^n}}\cdot\mathbf{\nabla}\rho\pm 2b\Big(\mathbf{\nabla}\cdot\textbf{\emph{v}}-S_{i}S_{j}\frac{\partial v_{i}}{\partial x_{j}}\Big)\Big] \\
&-2\frac{\big<z^{\infty\pm2}\big>\big<z^{\infty\mp2}\big>^\frac{1}{2}\big<z^{*\pm2}\big>}{L_\infty^\pm}-2\frac{\big<z^{*\mp2}\big>^\frac{1}{2}\big<z^{*\pm2}\big>^2}{L_*},
\end{aligned}
\end{equation}
where $b$ is a structural similarity parameter associated specifically with relating the cross-correlations of the magnetic field fluctuations to the 1-point magnetic field correlation~\citep{Zank12} which we take as $b=0.3$, and $\textbf{\emph{S}}$ is the slab direction defined by the mean magnetic field $\textbf{$\emph{B}_{0}$}$;
\begin{equation}
\label{eq12}
\begin{aligned}
\big(RHS\big)_{E_D^*} =
&\frac{1}{2}{E_D^*}\mathbf{\nabla}\cdot\textbf{\emph{v}}-\Big(3b-\frac{1}{2}\Big){E_T^*}\mathbf{\nabla}\cdot\textbf{\emph{v}}+3b{E_T^*}S_{i}S_{j}\frac{\partial v_{i}}{\partial x_{j}}+b{E_D^*}\Big(\mathbf{\nabla}\cdot\textbf{\emph{v}}-S_{i}S_{j}\frac{\partial v_{i}}{\partial x_{j}}\Big) \\
&+2b{E_C^*}\Big(\mathbf{\nabla}\cdot\textbf{$\emph{v}_{A}$}-S_{i}S_{j}\frac{\partial v_{A_i}}{\partial x_{j}}\Big)+\frac{1}{2}{E_C^*}\frac{1}{\rho}\textbf{$\emph{v}_{A}$}\cdot\mathbf{\nabla}\rho+\frac{1}{\rho}b{E_C^*}\Big(\textbf{$\emph{v}_{A}$}\cdot\mathbf{\nabla}\rho-S_{i}S_{j}\frac{\partial\rho}{\partial x_{j}}v_{A_i}\Big) \\
&+\frac{1}{4}\frac{1}{\rho}\Big({E_D^*}\big(\big<z^{\infty+2}\big>^\frac{1}{2}+\big<z^{\infty-2}\big>^\frac{1}{2}\big)-\big<z^{*-2}\big>\big<z^{\infty+2}\big>^\frac{1}{2}-\big<z^{*+2}\big>\big<z^{\infty-2}\big>^\frac{1}{2}\Big)\textbf{\emph{\^n}}\cdot\mathbf{\nabla}\rho \\
&-{E_D^*}\Big(\frac{\big<z^{\infty-2}\big>^\frac{1}{2}\big<z^{\infty+2}\big>}{L_\infty^+}+\frac{\big<z^{\infty+2}\big>^\frac{1}{2}\big<z^{\infty-2}\big>}{L_\infty^-}\Big)-{E_D^*}\Big(\frac{\big<z^{*+2}\big>^\frac{1}{2}\big<z^{*-2}\big>}{L_*}+\frac{\big<z^{*-2}\big>^\frac{1}{2}\big<z^{*+2}\big>}{L_*}\Big),
\end{aligned}
\end{equation}
where $E_T^*=\big(\big<z^{*+2}\big>+\big<z^{*-2}\big>\big)/2$ is the total energy in slab fluctuations, and $E_C^*=\big(\big<z^{*+2}\big>-\big<z^{*-2}\big>\big)/2$ is the slab cross-helicity;
\begin{equation}
\label{eq13}
\begin{aligned}
\big(RHS\big)_{L_*} =
&\frac{1}{2}{L_*}\mathbf{\nabla}\cdot\textbf{\emph{v}}-\Big(b-\frac{1}{4}\Big){L_D^*}\mathbf{\nabla}\cdot\textbf{\emph{v}}+b{L_D^*}S_{i}S_{j}\frac{\partial v_{i}}{\partial x_{j}} \\
&-\frac{1}{2}\Big({L_*}-\frac{L_D^*}{2}\Big)\Big[-\big<z^{\infty\pm2}\big>^\frac{1}{2}\frac{1}{\rho}\textbf{\emph{\^n}}\cdot\mathbf{\nabla}\rho-2b\Big(\mathbf{\nabla}\cdot\textbf{\emph{v}}-S_{i}S_{j}\frac{\partial v_{i}}{\partial x_{j}}\Big)\Big];
\end{aligned}
\end{equation}
and
\begin{equation}
\label{eq14}
\begin{aligned}
\big(RHS\big)_{L_D^*} = 
&\frac{1}{2}{L_D^*}\mathbf{\nabla}\cdot\textbf{\emph{v}}+b{L_D^*}\mathbf{\nabla}\cdot\textbf{\emph{v}}-\big(6b-1\big){L_*}\mathbf{\nabla}\cdot\textbf{\emph{v}}+6b{L_*}S_{i}S_{j}\frac{\partial v_{i}}{\partial x_{j}} \\
&-b{L_D^*}S_{i}S_{j}\frac{\partial v_{i}}{\partial x_{j}}-\frac{1}{2}\Big[\Big({L_*}-\frac{L_D^*}{2}\Big)\big<z^{\infty+2}\big>^\frac{1}{2}+\Big({L_*}-\frac{L_D^*}{2}\Big)\big<z^{\infty-2}\big>^\frac{1}{2}\Big]\frac{1}{\rho}\textbf{\emph{\^n}}\cdot\mathbf{\nabla}\rho.
\end{aligned}
\end{equation}

For the derivation of the NI MHD model equations, the interested reader can refer to~\citet{Zank17}.

\subsection{MHD Coronal Model}
\label{corona}

The governing equations that we solve to model the background coronal plasma in the loop are the system of ideal MHD equations. We write this system in differential, conservative form as follows:
\begin{equation}
\label{eq15}
\resizebox{.7\hsize}{!}{$\frac{\partial}{\partial t}
\left(
\begin{array}{c}
\rho  \\
\rho \textbf{\emph{v}} \\
\textbf{\emph{B}} \\
E
\end{array}
\right) +
\mathbf{\nabla}\cdot
\left(
\begin{array}{c}
\rho\textbf{\emph{v}} \\
\rho\textbf{\emph{v}}\textbf{\emph{v}} + \textbf{\emph{I}}(p+\frac{B^2}{8\pi})-\frac{\textbf{\emph{B}}\textbf{\emph{B}}}{4\pi} \\
\textbf{\emph{v}}\textbf{\emph{B}}-\textbf{\emph{B}}\textbf{\emph{v}} \\
(E+p+\frac{B^2}{8\pi})\textbf{\emph{v}}-\frac{\textbf{\emph{B}}}{4\pi}(\textbf{\emph{v}}\cdot\textbf{\emph{B}})
\end{array}
\right)
=
\left(
\begin{array}{c}
0 \\
0 \\
0 \\
S_{E}
\end{array}
\right),$}
\end{equation}
where $\textbf{\emph{I}}$ is the 3$\times$3 identity matrix, $\rho$, $\textbf{\emph{v}}$, $\textbf{\emph{B}}$, $p$, and $E$ are the density, velocity, magnetic field, thermal pressure, and specific total energy of the plasma, respectively. 

The specific total energy of the plasma, $E$, is given as follows:
\begin{equation}
\label{eq16}
E=\frac{p}{\gamma-1}+\frac{1}{2}\rho v^2+\frac{B^2}{8\pi},
\end{equation}
where $\gamma$ is the ratio of specific heats which we take as $\gamma=5/3$. The plasma is assumed to obey the ideal gas law and to be calorically perfect, which is a very good approximation for most space and solar plasmas. The ideal gas law together with Eq.~\ref{eq16} are necessary constitutive relations to close the set of ideal MHD equations. Finally, there is the solenoidal constraint ($\mathbf{\nabla}\cdot\textbf{\emph{B}}=0$) that should be satisfied, which can be recovered analytically from Eq.~\ref{eq15} by taking the divergence of the magnetic induction equation, supposing divergence free initial conditions.

$S_E$ is the coronal heating term due to MHD turbulence transported by the NI MHD turbulence transport model: 
\begin{equation}
\label{eq17}
\begin{aligned}
S_E = 
&\rho\alpha_{KT}\Big[2\frac{\big<z^{*+2}\big>\big<z^{\infty+2}\big>\big<z^{\infty-2}\big>^\frac{1}{2}}{L_\infty^+}+2\frac{\big<z^{*-2}\big>\big<z^{\infty-2}\big>\big<z^{\infty+2}\big>^\frac{1}{2}}{L_\infty^-}+2\frac{\big<z^{*-2}\big>^\frac{1}{2}\big<z^{*+2}\big>^2}{L_*} \\
&+2\frac{\big<z^{*+2}\big>^\frac{1}{2}\big<z^{*-2}\big>^2}{L_*}+2\frac{\big<z^{\infty+2}\big>^2\big<z^{\infty-2}\big>^\frac{1}{2}}{L_\infty^+}+2\frac{\big<z^{\infty-2}\big>^2\big<z^{\infty+2}\big>^\frac{1}{2}}{L_\infty^-}\Big],
\end{aligned}
\end{equation}
where $\alpha_{KT}$ is von K\'arm\'an-Taylor constant~\citep{Matthaeus96} which we take as $\alpha_{KT}=0.3$.

The NI MHD and ideal MHD systems of equations are coupled through the coronal heating term given in Eq.~\ref{eq17} and solved simultaneously at each iteration in a time-dependent fashion.

For more detailed information about our MHD coronal model, we refer the interested reader to~\citet{Yalim17,Singh18}.

\subsection{RMHD Model}
\label{RMHD}

The RMHD model describes the generation, propagation and dissipation of Alfv\'{e}n waves in a thin flux tube surrounding the axial guide magnetic field line. 
The tube has a circular cross-section with radius R(s) and starts from the base of the photosphere at one end, stretches through the chromosphere into the corona, and ends at the photosphere at the other end. 
The tube has a length L and we use a straightened tube, as is done commonly e.g., \citep{RP13}, i.e., the overall curvature of the tube is neglected. We use the coordinate system ${x,y,s}$, where $s$ is the coordinate along the flux tube axis $0\leq s \leq L$, and $x$ and $y$ are perpendicular to the loop axis.

The expansion factor of the field line is  $\Gamma \equiv B_{\rm TR} / B_{\rm min}$, where $B_{\rm min}$ is the minimum field strength in the corona and $B_{\rm TR}$ is the average of the field strengths at the two transition regions (TRs). The tube extends from the base of the photosphere at one end to the photosphere at the other end of the coronal loop. 

The background magnetic field strength $B_0(s)$ and plasma density $\rho_0(s)$ are functions of position $s$ only and are considered to be constant over the cross-section of the loop. Therefore, the Alfv\'{e}n
speed $v_A (s)$ ($\equiv B / \sqrt{4 \pi \rho}$) is also constant over the cross-section of the loop. The mass flows along the flux tube are neglected. The temperature $T_0(s)$ is a function of height and is based on a model of the lower atmosphere developed by \citet[]{Fontenla1999, Fontenla2006}. It is computed from

\begin{equation}
T_0(s) = T_{max} \left [ 1 - 0.8 u^2 (s) \right]^{2/7} ,
\end{equation}
where $u(s) \equiv -1 + 2(s-z_{TR})/L_{cor}$, which lies in the range $-1 \le u \le +1$, $z_{\rm TR}$ is the transition-region (TR) height, and $T_{max}$ is the peak temperature in the loop (in K) as predicted by the RTV scaling law \cite[][]{Rosner1978},

\begin{eqnarray}
T_{max} & \approx & 1.4 \times 10^3 (p_{cor} L_{cor} /2 )^{1/3} 
= 1.9 \times 10^6 ~ p_{cor}^{1/3} \left( \frac{L_{cor}}
{\mbox{50 Mm}} \right)^{1/3} ~~~~ {\rm K} , \label{eq:Tmax}
\label{eq:QcorRTV}
\end{eqnarray}
with $p_{cor}$ the
coronal plasma pressure (in $\rm dyne ~ cm^{-2}$), and $L_{cor}$ the coronal loop length (in cm or Mm).

In the photosphere, at the two ends of the flux tube ($s=0$ and $s=L$), we impose random footpoint motions. These footpoint motions consist of two counter-rotating cells with arbitrary orientation, and create transverse motions in the plasma along the magnetic field line. The Alfv\'{e}n waves produced as a result of these motions travel upward and propagate along the flux tube.
 The waves reflect due to the
spatial variations of Alfv\'{e}n speed $v_A(s)$.
The reflection of the waves at different heights produces
counter-propagating waves that interact with each other non-linearly
and produce Alfv\'{e}n wave turbulence. In our numerical calculation,
we start by assuming a  root-mean-square (rms) velocity of 1.48
$\rm km ~ s^{-1}$ for the footpoint motions and a correlation time of
$\tau_{c} = 60/\sqrt{2 \pi}=24$ s. 

In the RMHD model, the magnetic and velocity fluctuations are simulated but their effects on temperature and density are ignored. The magnetic field fluctuations ${\bf B}_1$ are considered to be small compared to the background field ($|{\bf B}_1 | \ll B_{0}$) and are computed as ${\bf B}_1 = \nabla_\perp h \times {\hat{\bf B_0}}$, where $h(x, y, s,t)$ is the magnetic flux function and $t$ is the time. The velocity fluctuations are assumed to be small compared to the Alfv\'{e}n speed $v_A(s)$. The velocity fluctuations are approximated by ${\bf v_1} = \nabla_\perp f \times {\hat{\bf B_0}}$, where $f(x, y, s, t)$ is the velocity stream function and $\hat {\bf B}_0 (x,y,s)$ is the unit vector along the background field, and $t$ is the time. The flows along the background field are neglected. The functions $f(x, y, s, t)$ and $h(x, y, s, t)$ satisfy the following equations:

\begin{equation}
\frac{\partial \omega} {\partial t} + \hat{\bf B}_0 \cdot
( \nabla_\perp \omega \times \nabla_\perp f ) = v_A^2 \left[
\hat{\bf B}_0 \cdot \nabla \alpha + \hat{\bf B}_0 \cdot
( \nabla_\perp \alpha \times \nabla_\perp h ) \right] + D_v , 
\label{eq:dodt}
\end{equation}
\begin{equation}
\frac{\partial h} {\partial t} = \hat{\bf B}_0 \cdot \nabla f
+ \frac{f} {H_B} + \hat{\bf B}_0 \cdot ( \nabla_\perp f \times
\nabla_\perp h ) + D_m ,
 \label{eq:dhdt}
\end{equation}
where $\omega$ ($\equiv - \nabla_\perp^2 f$) is the parallel component of vorticity, $\alpha$ ($\equiv -\nabla_\perp^2 h$) is the magnetic torsion parameter, $H_B (s) \equiv B_0/(dB_0/ds)$ is the magnetic scale length defined in Eq~\ref{eq:HB}. The terms $D_v$ and $D_m$ correspond to the effects of
viscosity and resistivity on the high wavenumber modes.

The kinetic and magnetic heating rates are defined as
\begin{equation}
Q_{kin} (s,t) \equiv \frac{\rho_0} {R^2} \sum_{k=1}^{N} \nu_k a_k^2
f_k^2, \label{eq:qkin}
\end{equation}
and
\begin{equation}
Q_{mag} (s,t) \equiv \frac{B_0} {4 \pi R^2} \sum_{k=1}^{N} \nu_k a_k^2
h_k^2, \label{eq:qmag}
\end{equation}
where $B_0$ is the background magnetic field strength, $\rho_0$ is the background density, $a_k$ is the perpendicular wavenumber, and $\nu_k$ is the damping rate. The waves are described in terms of their transverse nature using a spectral method presented in Appendix B of \citet[][]{van Ballegooijen11}. 

The total dissipation rate is $Q(s,t) \equiv Q_{kin} + Q_{mag}$. Derivations of the above equations and the detailed descriptions of their numerical implementation are given in \citet[][]{van Ballegooijen11}.

\section{Results} 
\label{res}

In this section, we present and discuss our results related to the numerical simulations that we performed to solve a realistic benchmark coronal loop heating problem by using the coupled MHD/NI MHD and RMHD models.

\subsection{MHD/NI MHD Model Simulation Setup and Results}
\label{sim}

We consider the loop geometry as a rectangular box (i.e., a straightened loop) in Cartesian coordinates where the loop axis coincides with the $z$-axis. Hence, the $z$ boundaries of the computational domain correspond to the footpoints of the loop which are located in the lower corona. The length of the loop is 48 Mm.

At t=0, the plasma inside the loop domain has a velocity of $\pm 30$ $\rm km ~ s^{-1}$ on both sides of the apex with a uniform axial guide magnetic field of $\textbf{$\emph{B}_{0}$}=100\textbf{\emph{\^k}}$ G where \textbf{\emph{\^k}} is the unit vector along the $z$-axis and constant density and temperature of $\rho_{0}=4.487\times10^{-15}$ $\rm g ~ cm^{-3}$ and $T_{0}=8.25\times10^5$ K that yield a thermal pressure of $p_{0} = 0.61$ $\rm dyne ~ cm^{-2}$. The initial conditions for the turbulence transport variables are uniform throughout the domain with values assigned from Table~\ref{tab1}.

\begin{figure}
\begin{centering}
\includegraphics[width=0.4\textwidth]{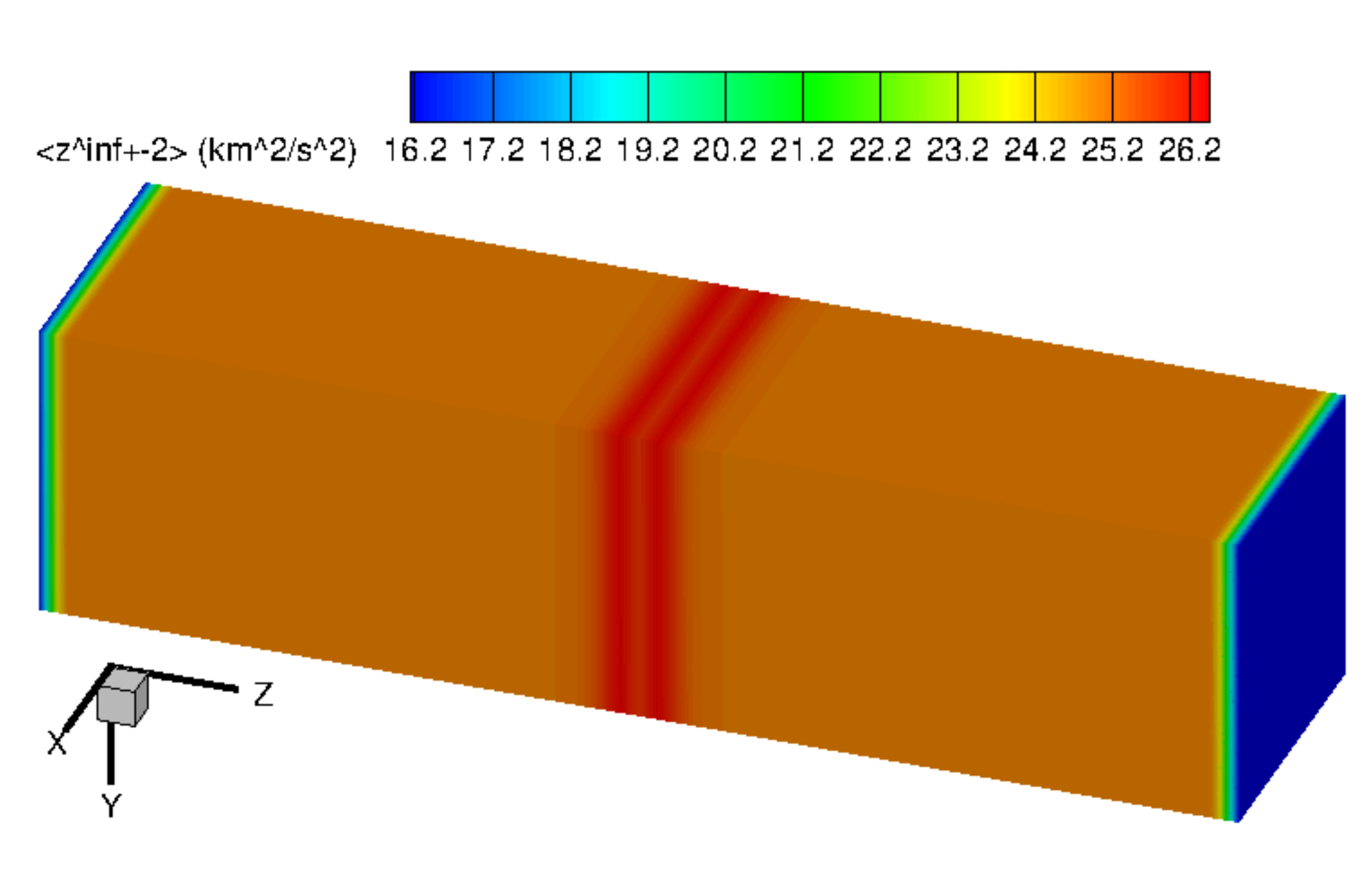}
\includegraphics[width=0.4\textwidth]{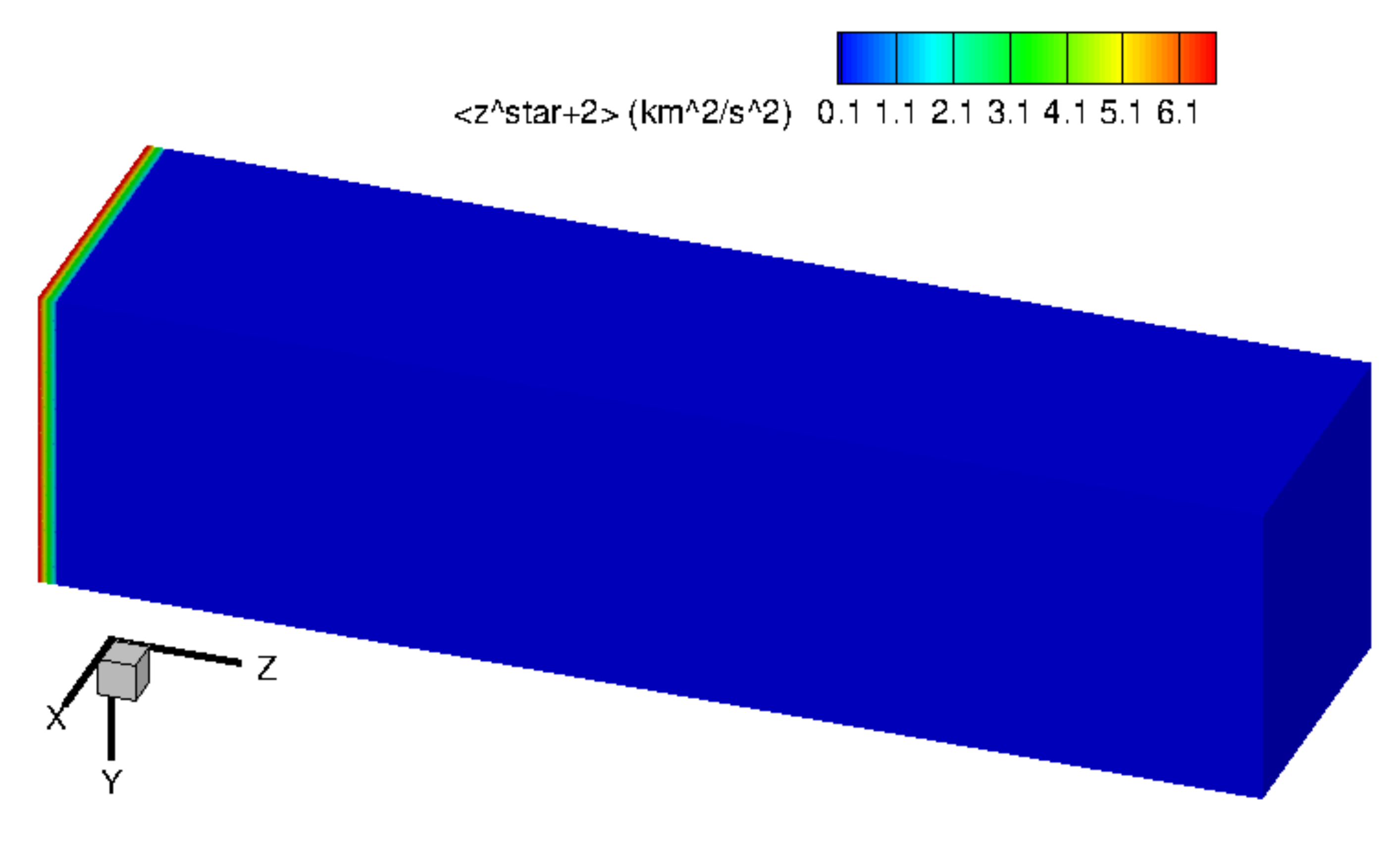}
\includegraphics[width=0.4\textwidth]{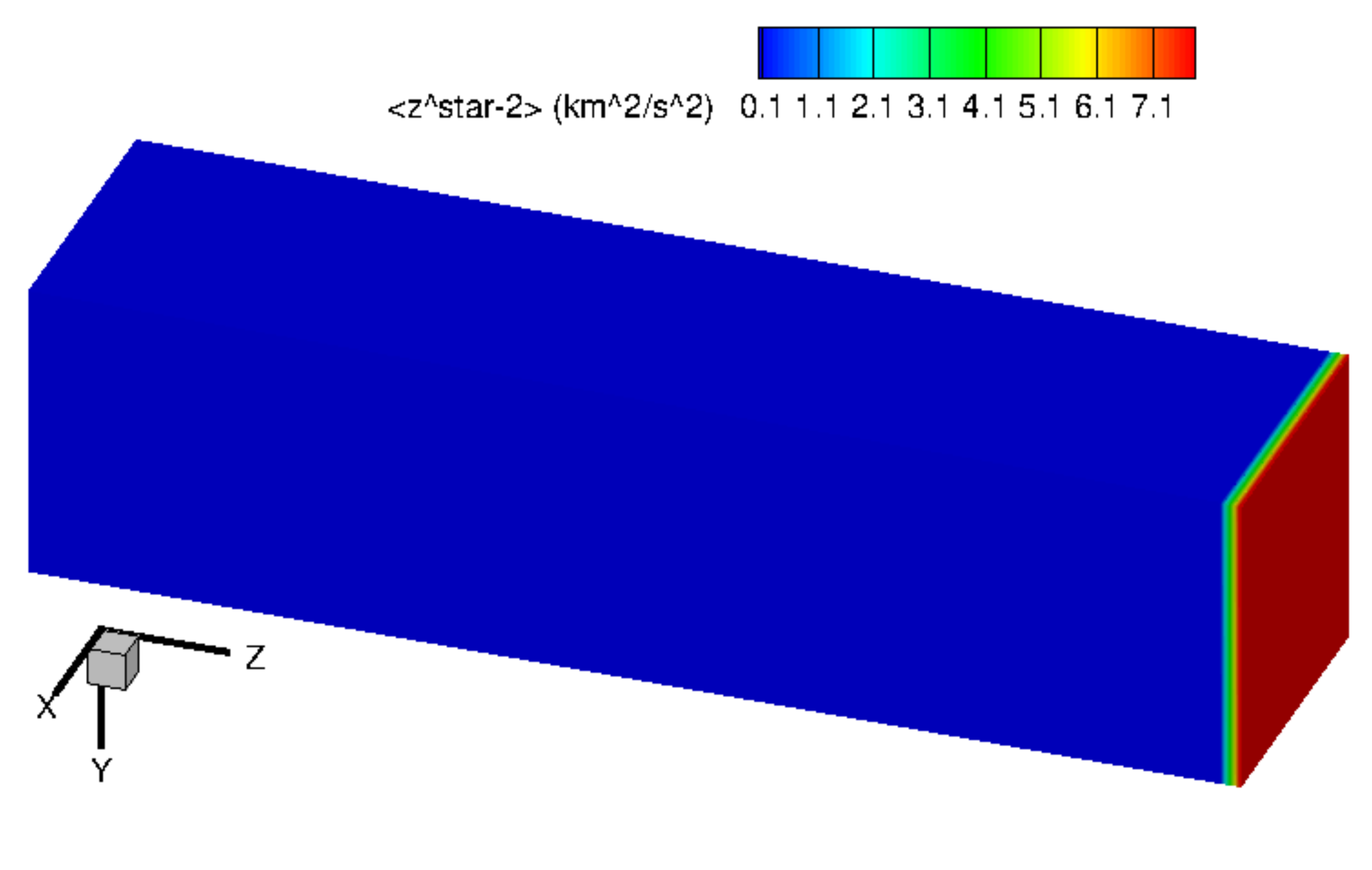}
\includegraphics[width=0.4\textwidth]{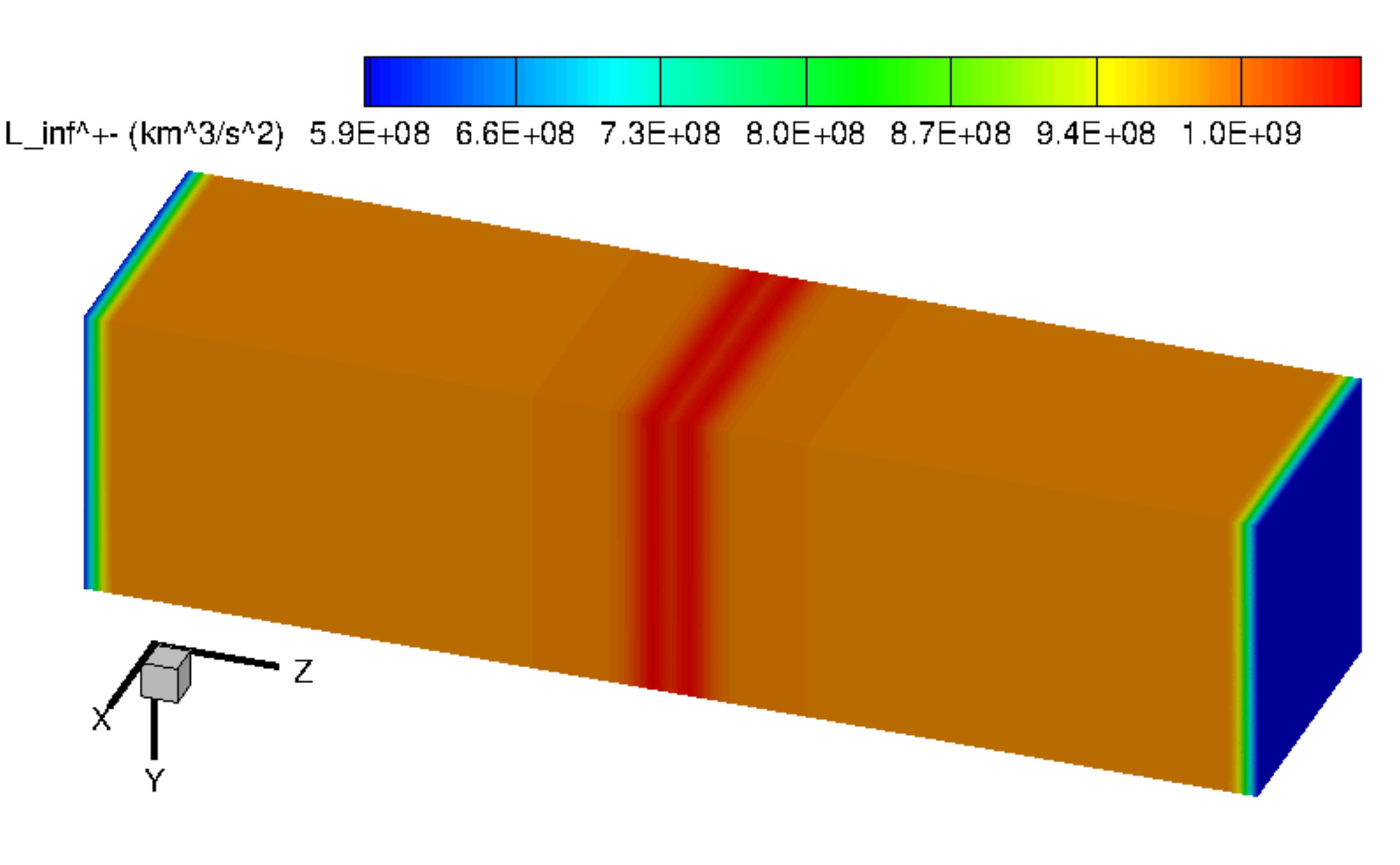}
\includegraphics[width=0.4\textwidth]{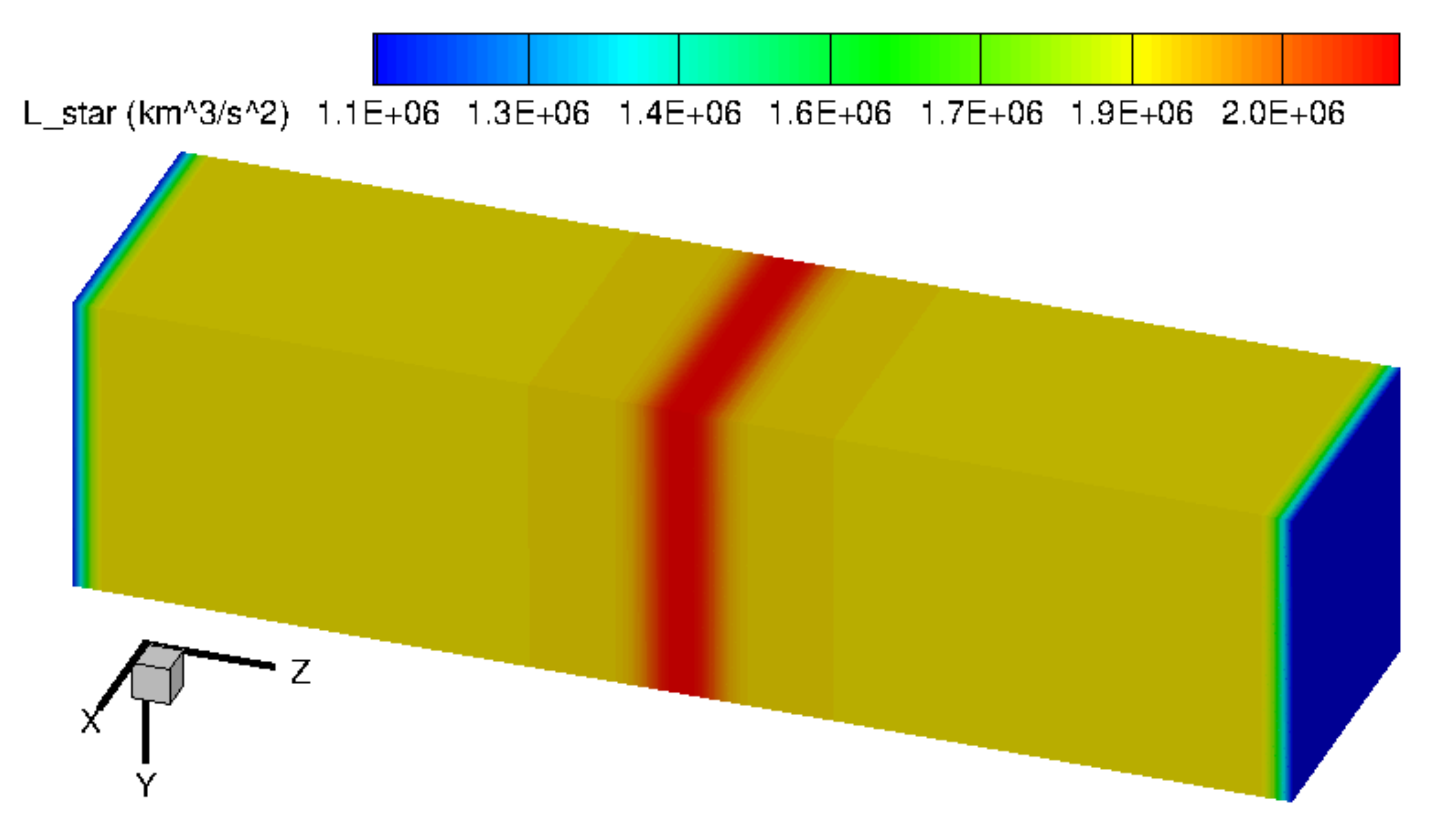}
\caption{Variations of the quasi-2D and slab Els\"asser variables ($\big<z^{\infty\pm 2}\big>$ and $\big<z^{*\pm 2}\big>$) and energy-weighted correlation lengths ($L_\infty^\pm$ and $L_*$) for backward/forward propagating modes (- and +) along the straightened coronal loop in the final solution at steady state}.
\label{figNIMHDSE}
\end{centering}
\end{figure}

At the $z$ boundaries, we impose an axial speed of 30 $\rm km ~ s^{-1}$ into the loop at both boundary surfaces for the turbulence fluctuations that are constantly imposed at the $z$ boundaries to be able to penetrate into the loop. Moreover, the gradients in magnetic field, density and specific total energy of the plasma are zero. The boundary values for the turbulence transport variables are tabulated again in Table~\ref{tab1}. The $x$ and $y$ boundaries are periodic.

\begin{table}[]
\centering
\resizebox{0.8\textwidth}{!}{
\begin{tabular}{|c|c|c|c|}
\hline
Quasi 2D variable & Value & Slab variable & Value \\ \hline
       $\big<z^{\infty\pm2}\big>$  & $2\times 10^{4}$\, $\textrm{km}^2/\textrm{s}^2$ & $\big<z^{*+2}\big>$ & $2.22222\times 10^{3}$\,$\textrm{km}^2/\textrm{s}^2$ \\ 
       $E_D^\infty$ & $-2.2\times 10^{3}$\,$\textrm{km}^2/\textrm{s}^2$ & $\big<z^{*-2}\big>$ & $5\times 10^{3}$\,$\textrm{km}^2/\textrm{s}^2$ \\ 
       $L_\infty^\pm$ & $1\times 10^{9}$\,$\textrm{km}^3/\textrm{s}^2$ & $E_D^*$ & $-1.1579\times 10^{2}$\,$\textrm{km}^2/\textrm{s}^2$ \\ 
      $L_D^\infty$ & $-1.1\times 10^{8}$\,$\textrm{km}^3/\textrm{s}^2$ & $L_*$ &  $1.92\times 10^{6}$\,$\textrm{km}^3/\textrm{s}^2$ \\ 
       $\big<\rho^{\infty2}\big>$ & $1.6\times 10^{45}$\,$\textrm{km}^{-6}$ & $L_D^*$ & $-2.89\times 10^{6}$\,$\textrm{km}^3/\textrm{s}^2$ \\ \hline

\end{tabular}}

\caption{Initial and boundary conditions for the NI MHD turbulence transport variables.}
\label{tab1}
\end{table}

The simulation was performed using the Multi-Scale Fluid-Kinetic Simulation Suite (MS-FLUKSS) code~\citep{Pogorelov14}. We utilize a cell-centered upwind Finite Volume method with ghost cells at the boundaries to spatially discretize the ideal MHD and NI MHD systems of equations. These equations are discretized in time using explicit schemes. More specifically, we apply the total variation diminishing (TVD) Roe's scheme and Hancock scheme to discretize the ideal MHD equations in space and time, and a TVD Courant-Isaacson-Rees scheme and Hancock scheme to discretize the NI MHD equations in space and time, respectively~\citep{Kryukov12}. Finally, the solenoidal constraint is satisfied using Powell's source term method~\citep{Powell99}.

Figure~\ref{figNIMHDSE} shows the variations of the quasi-2D and slab Els\"asser variables and energy-weighted correlation lengths for backward/forward propagating modes, namely $\big<z^{\infty\pm 2}\big>$, $\big<z^{*\pm 2}\big>$, $L_\infty^\pm$, and $L_*$, respectively, along the loop in the final solution at steady state. These turbulence transport variables are used together with the density, obtained from the corona model, to calculate the coronal heating term given in Eq.~\ref{eq17}. All these transport variables, especially the 2D and slab Els\"asser variables, decrease significantly from their initial values given in Table~\ref{tab1} (i.e., both by three orders of magnitude) resulting in the largest heating rate occurring at the starting time which decreases with time. This result shows that the majority 2D component plays a role as important as that of the minority slab component in heating the coronal loop. Results related to the plasma variables and magnetic field together with the heating rate are shown in Figure~\ref{figNIMHDRMHDresults} in subsection~\ref{heat}.

\subsection{RMHD Model Simulation Setup and Results}

\begin{figure*}
\begin{centering}
\includegraphics[width=4.00in]{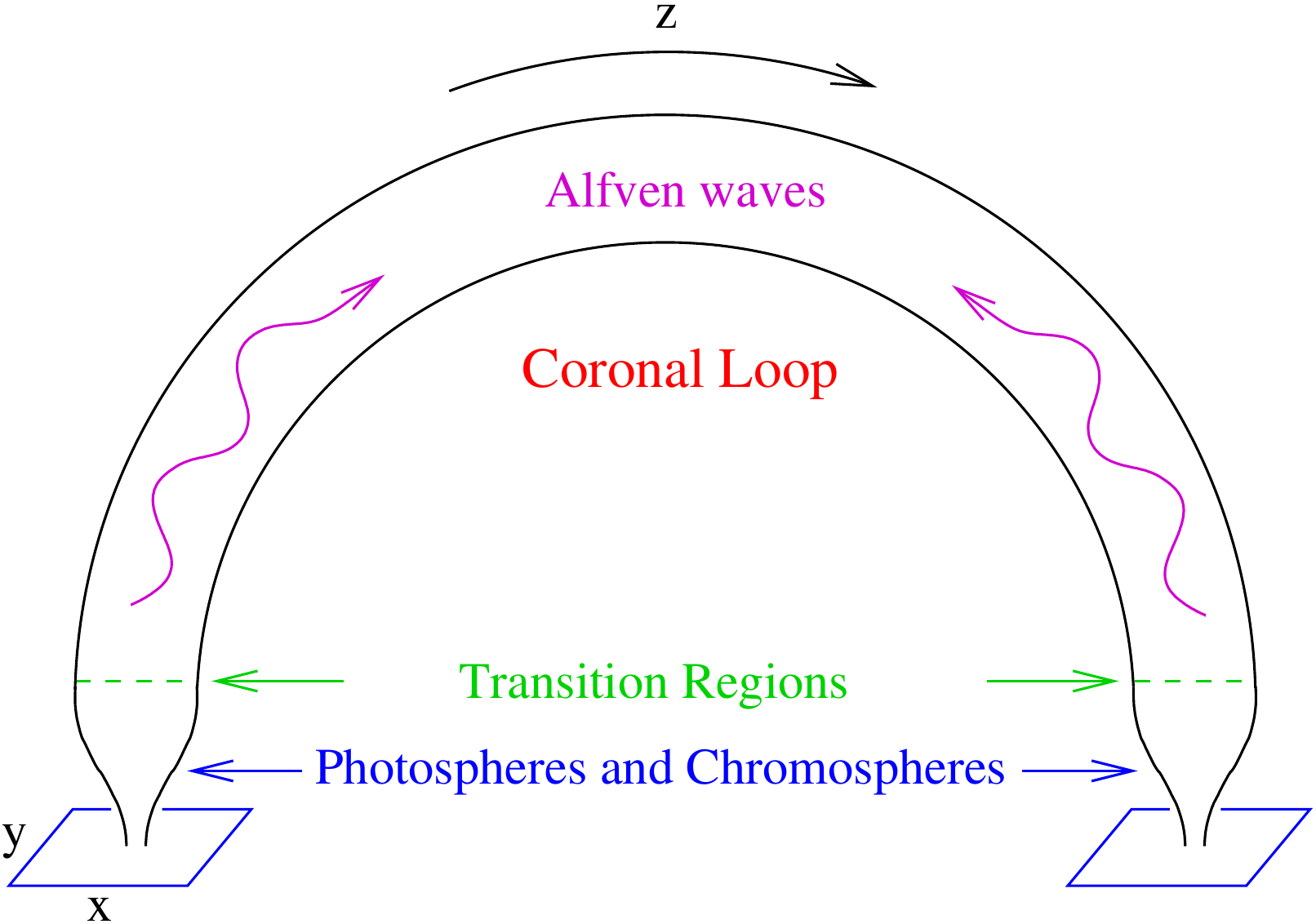}
\caption{Model for Alfv\'{e}n wave turbulence in coronal loops. The Alfv\'{e}n waves are driven by foot-point motions inside the tube. Note that in the RMHD approximation, the coronal loop is approximated with a straightened magnetic flux tube.}
\label{figcarl}
\end{centering}
\end{figure*}

We construct a model with coronal field strength
$B_{cor} = 100$ G, expansion factor $\Gamma = 1$, and coronal loop length of $L_{cor} = 48$ Mm, and the transition-region (TR) height $z_{TR} = 1.8$ Mm. The coronal loop footpoints are on the photosphere as shown in Figure~\ref{figcarl} and their motions have a correlation time $\tau_0 = 60$ s, each of the driver modes has a vorticity $\omega_0 = 0.04$ $\rm s^{-1}$, and the rms velocity is $\Delta v_{rms} = 1.48$ $\rm km ~ s^{-1}$. The TR height corresponds to a coronal pressure $p_{cor} = 0.61$ $\rm dyne ~ cm^{-2}$, which is typical for some of the warm loops found in active regions, and yields a peak temperature $T_{max} = 1.59$ MK.

\begin{figure*}
\begin{centering}
\includegraphics[width=0.4\textwidth]{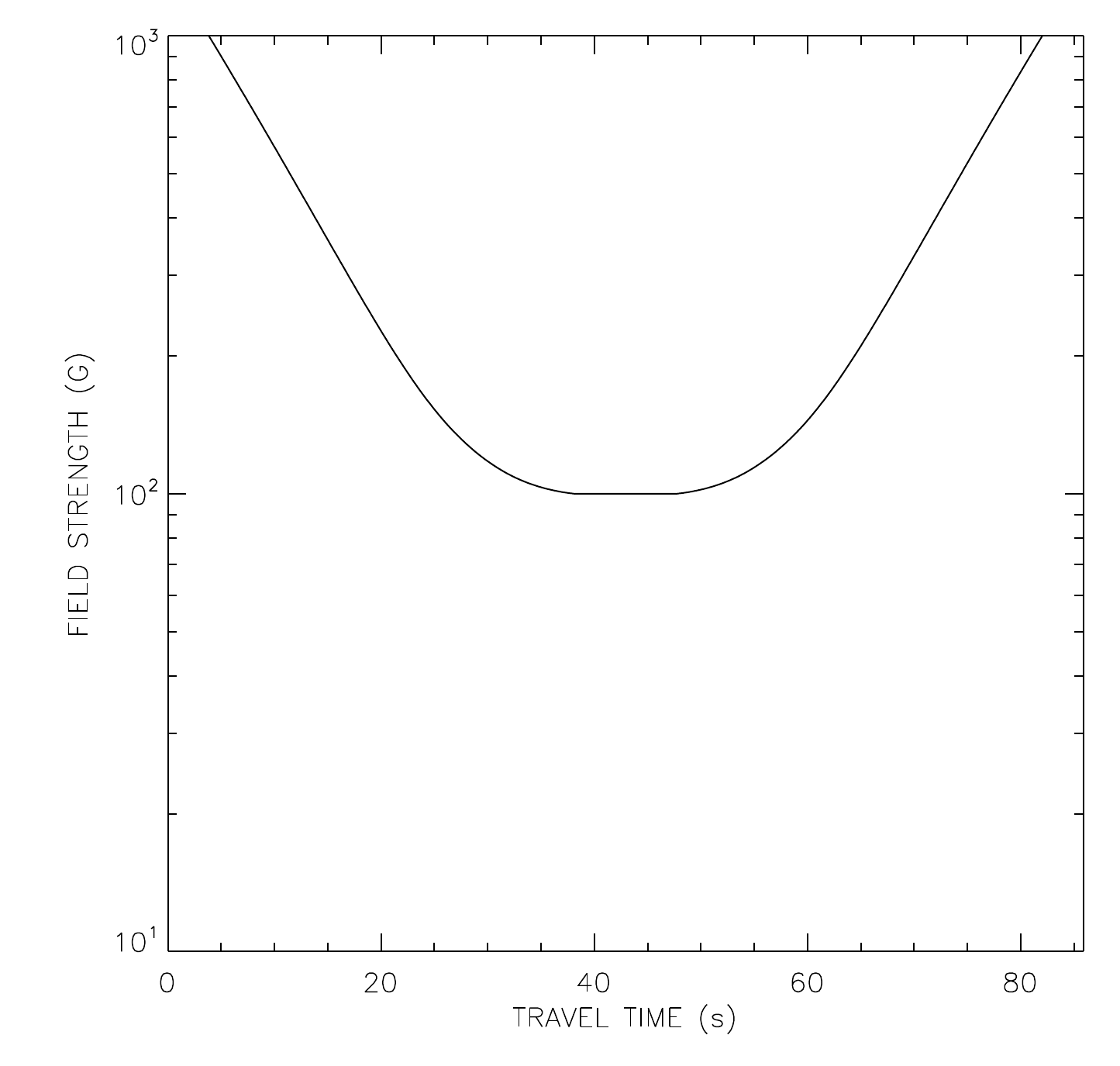}
\includegraphics[width=0.4\textwidth]{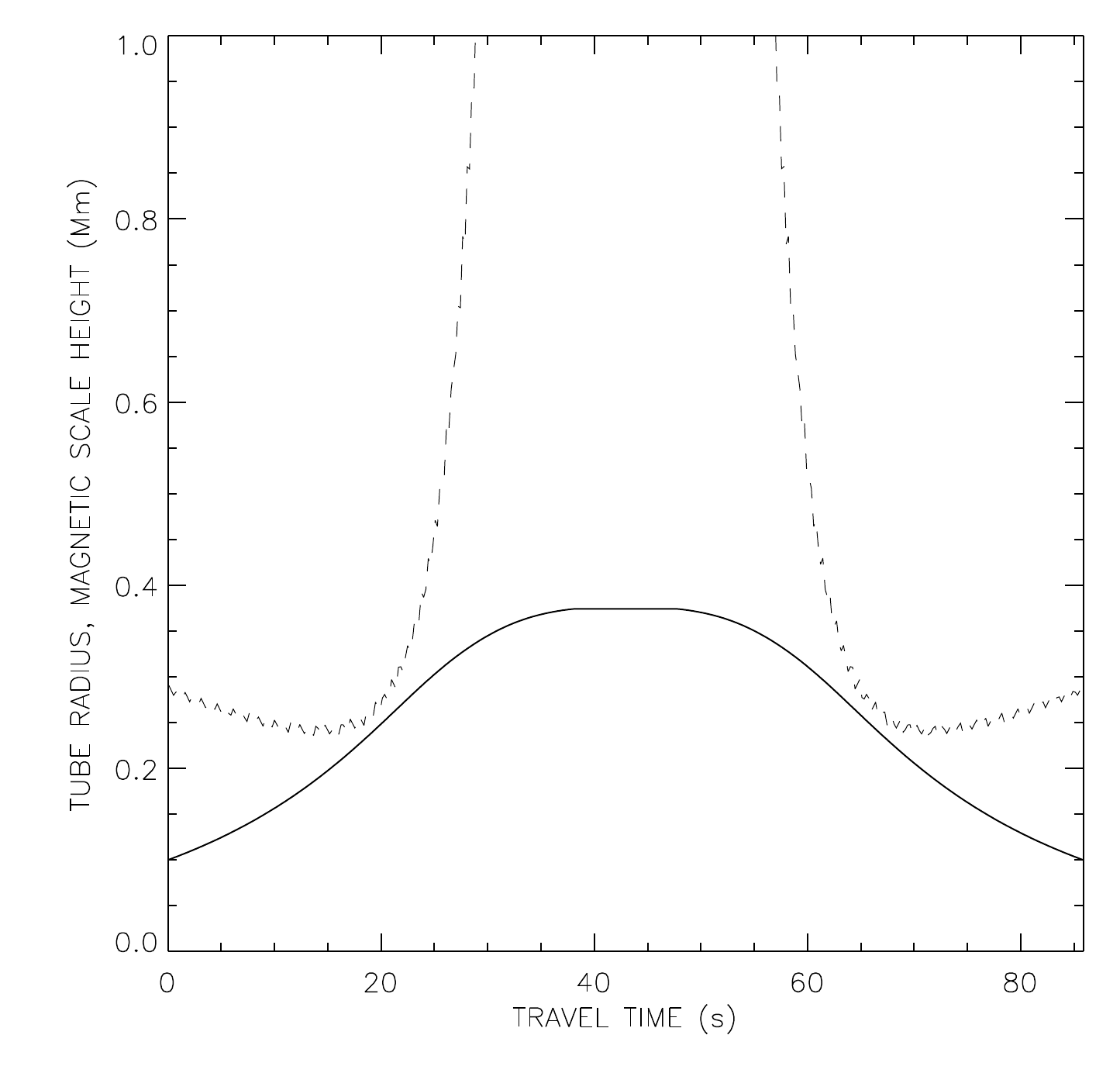}
\caption{(a) The magnetic field strength $B_0$, (b) the flux tube radius $R$ ({\it full} curve) and the magnetic scale height $| H_B |$ ({\it dashed} curve).}
\label{TwoPanels}
\end{centering}
\end{figure*}

The background field ${\bf B}_0$ is non-uniform and varies on a spatial
scale $H_B$, which is defined by
\begin{equation}
H_B \equiv B_0 \left( \hat{\bf B}_0 \cdot \nabla B_0 \right)^{-1} ,
\label{eq:HB}
\end{equation}
where $B_0 ({\bf r})$ is the background field strength, and
$\hat{\bf B}_0 ({\bf r})$ is the unit vector along the background
field. Figure~\ref{TwoPanels} shows the magnetic field strength $B_0$, the flux tube radius $R$ ({\it full} curve) and the magnetic scale height $| H_B |$ ({\it dashed} curve). 

\begin{figure*}
\begin{centering}
\includegraphics[width=6.00in]{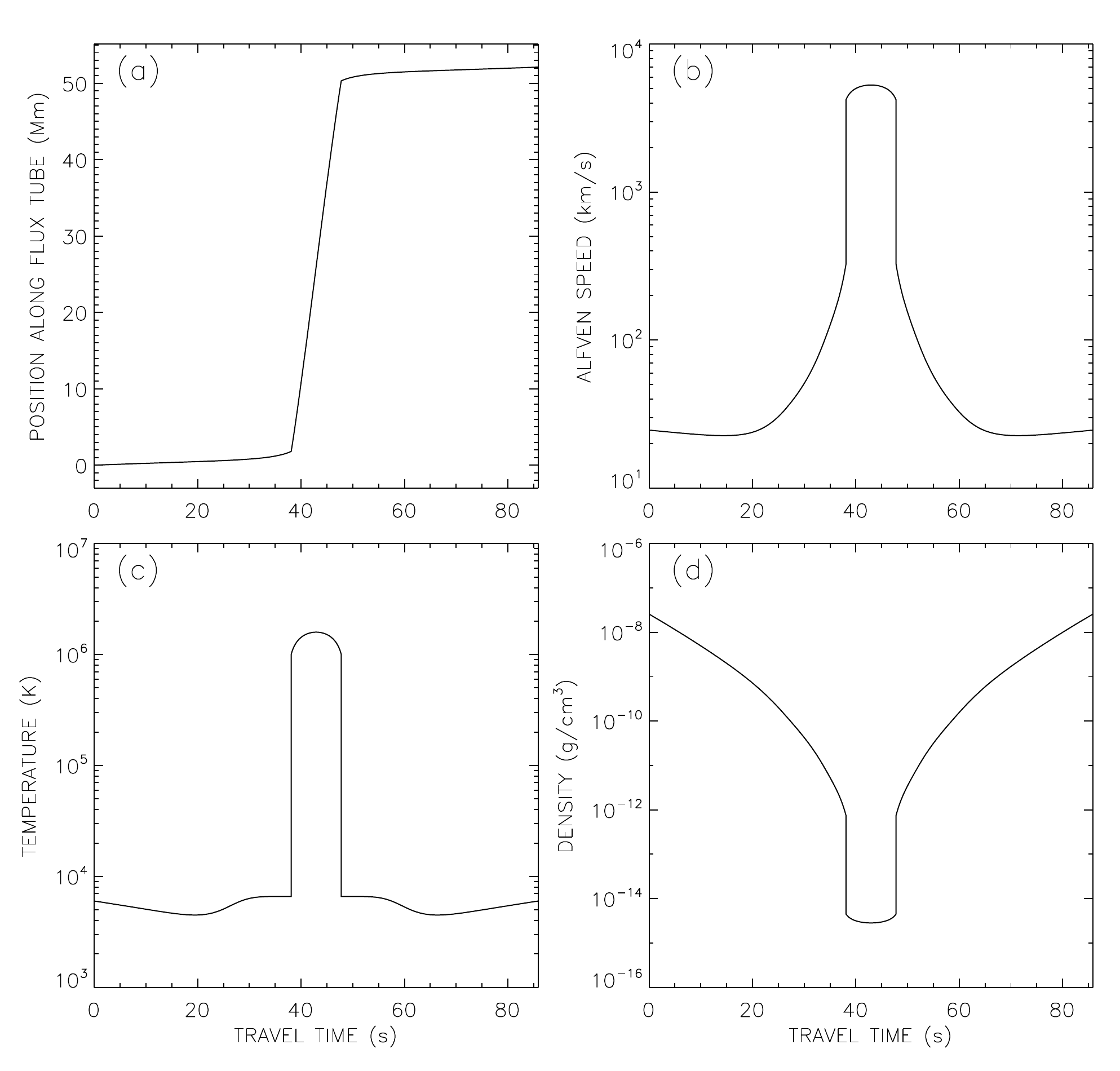}
\caption{Various quantities are plotted as a function of the Alfv\'{e}n wave travel time $\tau$: (a) position $s(\tau)$ along the loop measured from the left footpoint, (b) Alfv\'{e}n speed $v_A$; (c) temperature $T_0$; and (d) mass density $\rho_0$. The two chromosphere-corona TRs are located at $\tau$ = 38.1 s and $\tau$ = 47.7 s.}
\label{figbck}
\end{centering}
\end{figure*}

Figure~\ref{figbck} shows various quantities
plotted as a function of position along the flux tube for this model. Positions are given in terms of the Alfv\'{e}n wave travel time from the left footpoint ($s=0$). Figure~\ref{figbck}(a) shows the relationship between s and $\tau$. The photospheric footpoints are located at
$\tau(0) = 0$ and $\tau(L) = 52.0$ s, and the corona is located in
the region 38.1 s $<$ $\tau$ $<$ 47.7 s. The other panels in the Figure
show the Alfv\'{e}n speed $v_A$, temperature $T_0$, and density $\rho_0$.

\begin{figure*}
\begin{centering}
\includegraphics[width=3.00in]{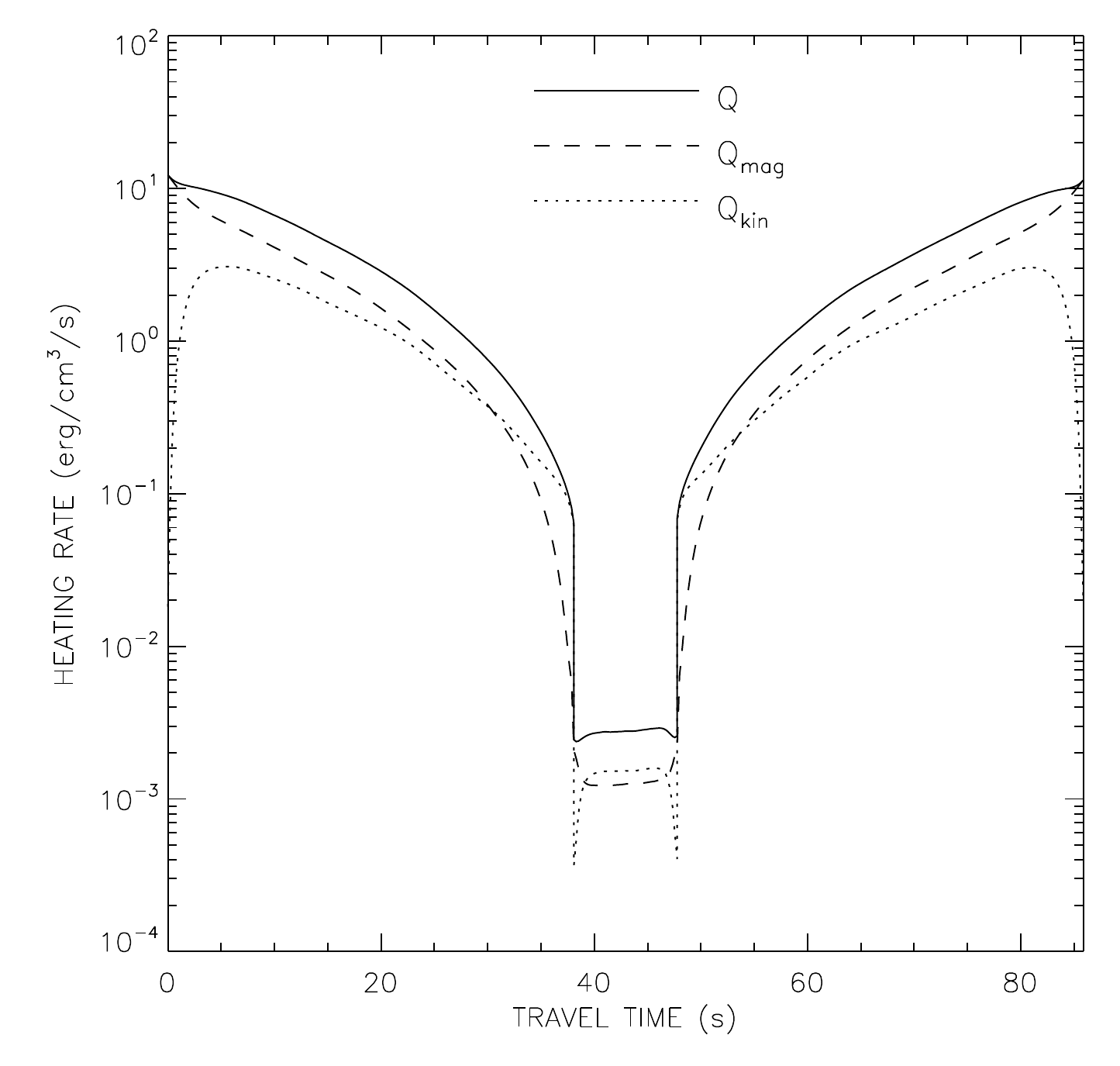}
\caption{Kinetic and magnetic heating
rates, and their sum $Q(s)$ as a function of Alfv\'{e}n travel time.}
\label{RMHDresults}
\end{centering}
\end{figure*}

The length of the simulation is $t_{\rm max} = 3000$ s, which is much longer than the Alfv\'{e}n
wave travel time along the entire loop ($\sim$200 s).

Figure \ref{RMHDresults} shows the heating rates as a function of position along the flux tube, averaged over the cross-section of the flux tube ($x$ and $y$) and over the time interval $t=[800,3000]$ s.
Position is given in terms of the Alfv\'{e}n
travel time $\tau (s)$ in seconds. The Figure shows the kinetic and magnetic heating rates, $Q_{kin}(s)$ and $Q_{mag}(s)$, and their sum $Q(s)$.  These quantities are discontinuous at the TR. Between the photospheric footpoints and the transition region ($\tau < 38.1$ s and
$\tau > 47.7$ s) and in the corona ($38.1 < \tau < 47.7$ s) the magnetic heating dominates, but in the chromosphere $Q_{kin} > Q_{mag}$.

\subsection{Comparison of MHD/NI MHD and RMHD Model Simulation Results}
\label{heat}

\begin{figure}
\begin{centering}
\includegraphics[width=0.4\textwidth]{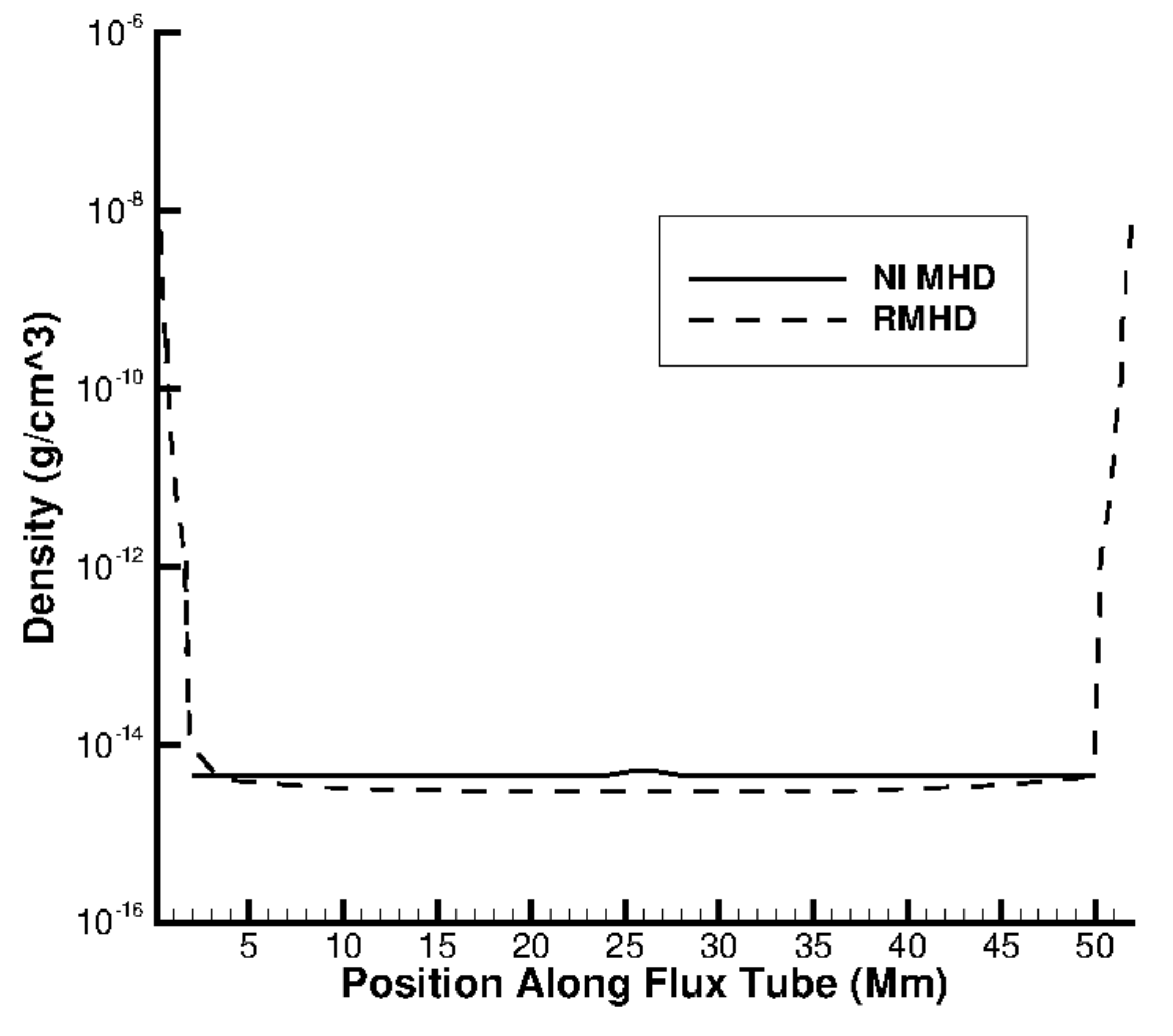}
\includegraphics[width=0.4\textwidth]{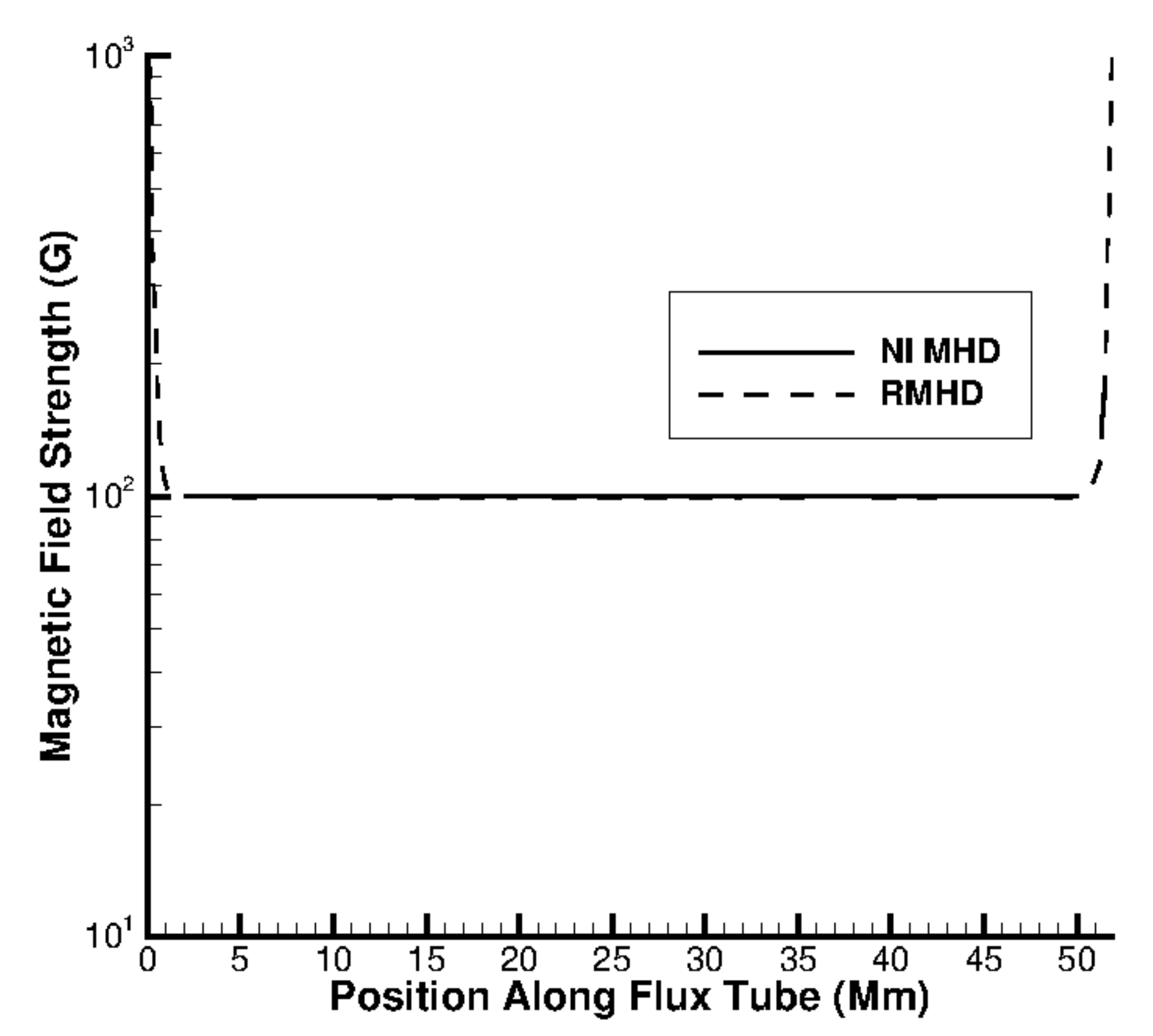}
\includegraphics[width=0.4\textwidth]{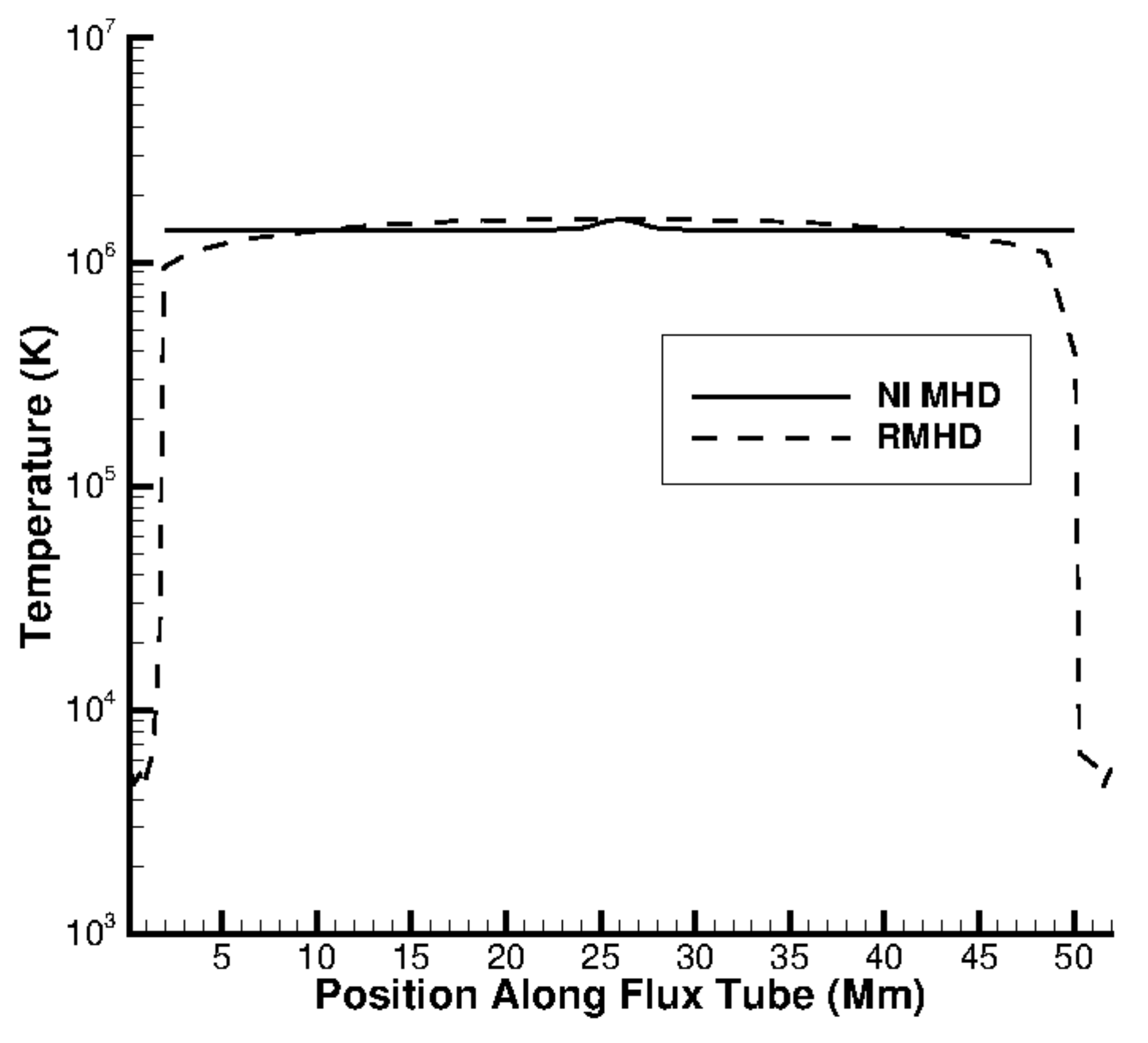}
\includegraphics[width=0.4\textwidth]{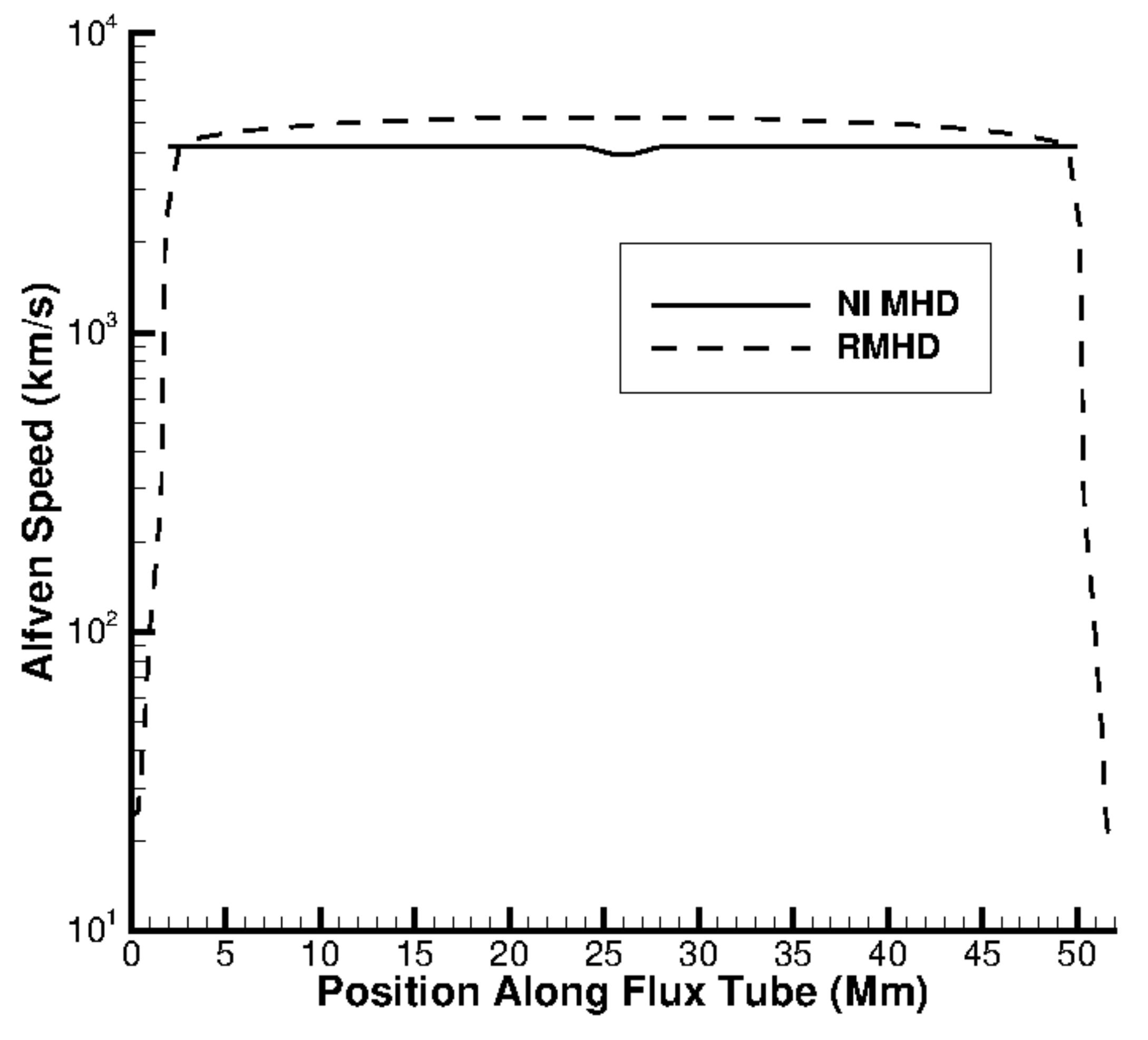}
\includegraphics[width=0.4\textwidth]{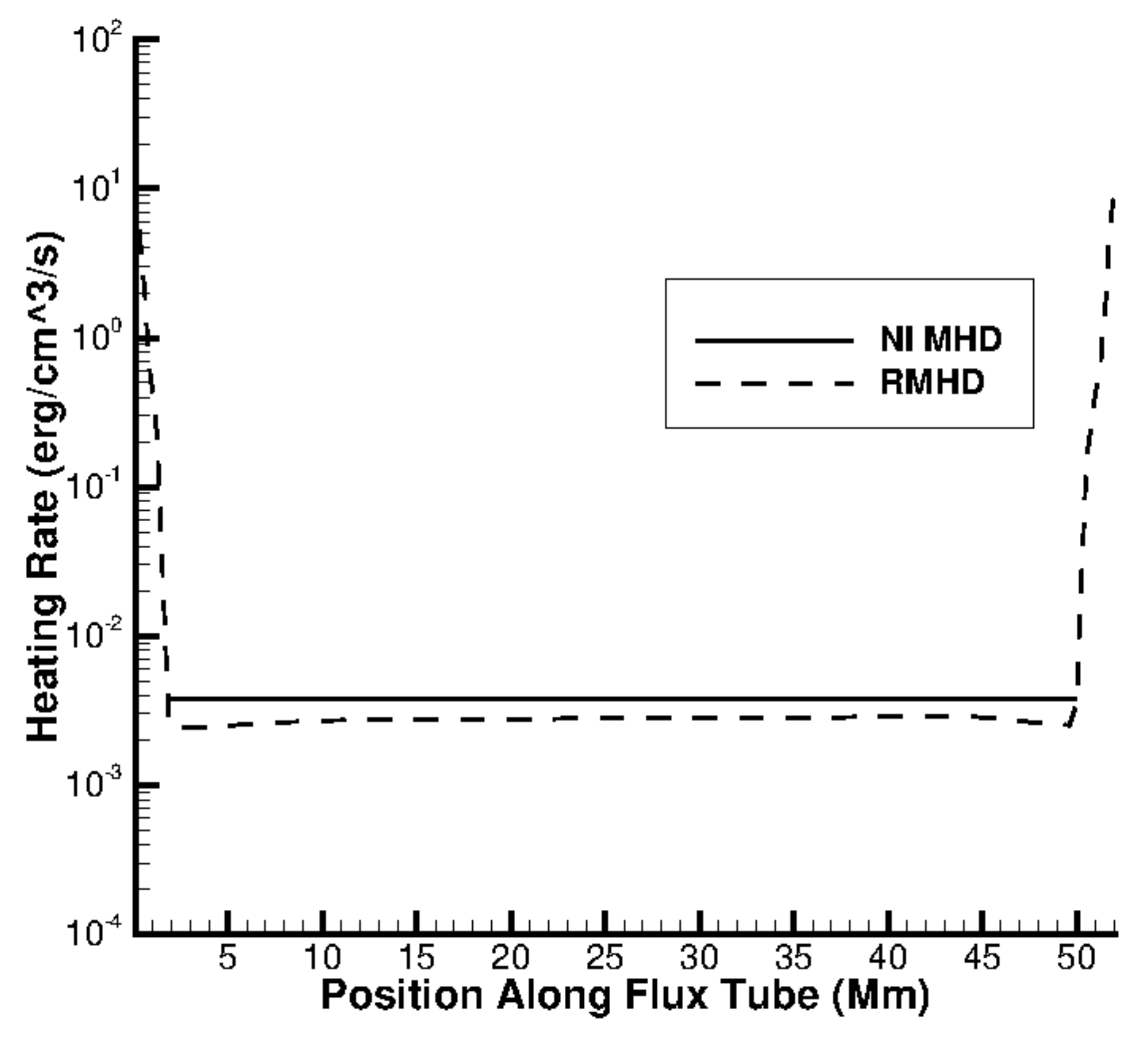}
\caption{NI MHD and RMHD model simulation results for the benchmark coronal loop heating problem: (Top row) (left) Density and (right) magnetic field strength along the loop; (Middle row) (left) temperature and (right) Alfv\'en wave speed along the loop; (Bottom row) coronal loop heating rate. The RMHD model simulation computational domain boundary is at the photosphere whereas it is at the lower corona for the NI MHD model simulation.}
\label{figNIMHDRMHDresults}
\end{centering}
\end{figure}

Based on the benchmark coronal loop heating problem that we simulated above using the coupled MHD corona/NI MHD turbulence transport model and the RMHD model,  we compare here the corresponding model results.

Figure~\ref{figNIMHDRMHDresults} shows the variations of density, magnetic field strength, temperature, Alfv\'en wave speed, and the coronal loop heating rate along the loop for both models. 
The density, magnetic field strength, temperature, and Alfv\'en wave speed are initial conditions for the RMHD model. These quantities are spatially averaged over the cross-section. The heating rate is calculated from the time-dependent RMHD model simulation which is also time-averaged in addition to being spatially averaged over the cross-section. For the NI MHD model results, all quantities are calculated from the time-dependent MHD/NI MHD model simulation and the heating rate is also time-averaged similar to the heating rate result from the RMHD model. Both model results show remarkably good agreement despite the basic differences in the approach which we will discuss below even if, at a very fundamental level, they derive from related physics.

The mechanism of how MHD turbulence is generated in both models is different. Within the confines of the RMHD model, itself containing certain assumptions that are elaborated above, the small-scale velocity and magnetic field fluctuations emerge directly from the simulation itself and are then dissipated via viscous and resistive dissipation. By contrast, the NI MHD model uses a mean field decomposition of the basic 3D time-dependent MHD equations and then certain closures for the fluctuations based on 1-point correlations to derive a set of evolution equations that describe the evolving energy-weighted correlation lengths. The energy-weighted correlation lengths can be interpreted in terms of total energy, residual energy, and cross helicity, and the system is closed by assuming that cross-correlations can be approximated by 1-point correlations via parameters $a$ and $b$. The dissipation of the fluctuations is based on the idea that the turbulence is fully developed and is described by a Kolmogorov (or if one wished an Iroshnikov-Kraichnan) phenomenology. Such a phenomenology allows one to ``short-circuit'' the details of the dissipation process, recognizing instead that the balancing of the energy input and the dissipation rate determines the (self-similar) cascade rate. Thus, the simulation in the NI MHD model solves two coupled systems of equations, one describing the large-scale background MHD flow (the MHD equations that have a heating term associated with the dissipation of turbulence) and the other being a turbulence transport model that includes the dissipation of the turbulence energy, described phenomenologically by the Kolmogorov model, as the turbulence is advected through the loop. Hence, in the NI MHD model, no small-scale fluctuations are introduced via the simulation unlike the RMHD model.

The computational domain boundary for the RMHD model simulation starts at the photosphere. In this model, Alfv\'en waves are generated by the footpoint motions of the coronal loop. While propagating along the loop, the generated waves travel forward and backward along the loop and interact with each other due to the flow gradients generating counter-propagating Alfv\'en waves that couple nonlinearly to produce turbulence. As shown in Figure~\ref{figbck}(b), the Alfv\'en wave speeds are two orders of magnitude smaller in the denser chromosphere and TR in comparison with their values in the corona, which can also be seen in the position along the loop vs. Alfv\'en wave travel time graph presented in Figure~\ref{figbck}(a). Additionally, braiding of magnetic field lines as well as small scale variations in transverse magnetic field and velocity inside the loop exist in the initial solution of the RMHD model. The turbulent relaxation of braided magnetic field structures plays a fundamental role in the loop heating mechanism described by the RMHD model.

In the NI MHD turbulence transport model, turbulence transport and evolution is solved directly from the model equations (derived from the MHD equations themselves via mean-field theory, suitable closures, and scale separation) and coronal heating, approximated using a phenomenological dissipation model of MHD turbulence, is expressed in terms of the transport variables. For the NI MHD turbulence transport model simulation, we simulated only the coronal part of the loop and did not impose any braiding. However, small-scale velocity or magnetic field fluctuations are present and evolved using the turbulence transport equations.

To compare the results between the NI MHD and RMHD results, since we cannot impose the loop footpoint boundary conditions at the same location for both models, we impose an initial solution based on the boundary conditions on the photosphere for the RMHD model simulation in a way that we can match the solution at the coronal footpoint boundaries of the NI MHD model.

At this point, we focus specifically on our comparison corresponding to the coronal loop heating rate presented in the bottom panel of Figure~\ref{figNIMHDRMHDresults} since this quantity is calculated from the time-dependent MHD/NI MHD and RMHD model simulation results. For the RMHD model, the heating rate is calculated from $Q$ which is the sum of $Q_{kin}$ and $Q_{mag}$ (see Eqs.~\ref{eq:qkin} and ~\ref{eq:qmag}). $Q_{kin}$ is the rate of kinetic energy loss due to damping and $Q_{mag}$ is the rate of magnetic energy loss. For the NI MHD model, the heating rate is calculated from the heating/decay phenomenology given by Eq.~\ref{eq17} which is a function of the turbulence transport variables corresponding to quasi-2D and slab Els\"asser variables and energy-weighted correlation lengths for backward/forward propagating modes. Despite the fundamental differences in the way the heating rate is calculated by both models, we obtain very good agreement between the time-averaged heating rates along the loop, which results in very similar temperature profiles with almost identical maximum temperature values at the apex of the loop (i.e., $T_{max}=1.54\times10^6$ K from the NI MHD model vs. $T_{max}=1.59\times10^6$ K from the RMHD model). We would like to emphasize here that the density, magnetic field strength, temperature, and Alfv\'en wave speed distributions given in Figure~\ref{figNIMHDRMHDresults} are part of the initial solution for the RMHD model while they were solved in a time-dependent fashion by the coupled MHD/NI MHD model simulation and correspond to the MHD/NI MHD model simulation results at the steady state.
 
\section{Conclusions} 
\label{conc}

In this paper, we used a benchmark problem to compare results from two different coronal heating models that are based on the transport of MHD turbulence within a realistic coronal loop setting. For our NI MHD turbulence transport model simulation, we simulated only the coronal part of the loop and did not impose any braiding. However, small-scale velocity or magnetic field fluctuations are present and evolved using the NI MHD model equations. The transport and dissipation of MHD turbulence was solved directly from the NI MHD model transport equations and coronal heating was expressed in terms of the transport variables. We found that the majority 2D component is as important as the minority slab component in the heating of the coronal loop. Our RMHD model simulation started from the photosphere. Alfv\'en wave turbulence was imposed by the footpoint motions of the loop in the presence of braiding. We imposed boundary conditions on the photosphere in a way that allowed us to match the solution at the coronal footpoints for both models. We also imposed the density, magnetic field strength, temperature and Alfv\'en wave speed as initial conditions for the RMHD model. We set these initial values according to the background coronal plasma solution of the coupled MHD/NI MHD model at the steady state. Despite the basic differences between the two models, the two sets of simulation results matched remarkably well, yielding almost identical heating rates inside the corona. This agreement between the NI MHD and RMHD model results is a very encouraging outcome of this work for the solar atmosphere modeling and coronal heating communities and demonstrates the importance of studies involving model comparisons. In future work, we will include model comparisons within coronal loops and in open magnetic field line regions based on solar observations. \\
 
\noindent
We thank the referee for their valuable comments and suggestions that improved the quality of our manuscript. We acknowledge support from the NSF EPSCoR RII-Track-1 Cooperative Agreement OIA-2148653. Any opinions, findings, and conclusions or recommendations expressed in this material are those of the author(s) and do not necessarily reflect the views of the National Science Foundation. We acknowledge the partial support of a Parker Solar Probe contract SV4-84017, M.S.Y. acknowledges partial support from NASA LWS grant 80NSSC19K0075 and NSF award AGS-2020703. M. A. T is supported by NASA contract NNM07AB07C (NASA Solar-B X-Ray Telescope Phase-E). We also acknowledge the Texas Advanced Computing Center (TACC) at The University of Texas at Austin for providing HPC resources that have contributed to the research results reported within this paper (URL: http://www.tacc.utexas.edu). We would like to thank Dr. Laxman Adhikari from The University of Alabama in Huntsville for discussions about the implementation of the NI MHD turbulence transport model.


\begin{thebibliography}{}
\bibitem[Adhikari et al.(2020)]{Adhikari20} Adhikari, L., Zank, G.~P., \& Zhao, L.-L.\ 2020, ApJ, 901, 102
\bibitem[Adhikari et al.(2021)]{Adhikari21} Adhikari, L., Zank, G.~P., Zhao, L.-L., et al. \ 2021, A\&A, 650, A16
\bibitem[Asgari-Targhi \& van Ballegooijen(2012)]{AvB12} Asgari-Targhi, M., \& van Ballegooijen, A.~A.\ 2012, ApJ, 746, 81
\bibitem[Asgari-Targhi et al.(2014)]{Asgari2014} Asgari-Targhi, M.,  van Ballegooijen, A.~A., \& Imada, S. 2014, ApJ, 786, 28
\bibitem[Asgari-Targhi et al.(2021)]{Asgari2021} Asgari-Targhi, M., Asgari-Targhi, A., Hahn, M., \& Savin, D.~W. 2021, ApJ, 911, 63
\bibitem[Chae, Sch\"{u}hle \& Lemaire(1998)]{Chae1998} Chae, J., Sch\"{u}hle, U., \& Lemaire, Ph. 1998, ApJ, 505, 957
\bibitem[Cranmer \& van Ballegooijen(2010)]{CvB10} Cranmer, S.~R., \& van Ballegooijen, A.~A. 2010, ApJ, 720, 824
\bibitem[Cranmer et al.(2015)]{Cranmer15} Cranmer, S.~R., Asgari-Targhi, M., Miralles, M.~P., et al. \ 2015, Philos. Trans. A Math. Phys. Eng. Sci., 373, 20140148
\bibitem[Dahlburg et al.(2012)]{Dahlburg12} Dahlburg, R.~B., Einaudi, G., Rappazzo, A.~F., et al. \ 2012, A\&A, 544, L20
\bibitem[Dere \& Mason(1993)]{Dere1993} Dere, K. P., \& Mason, H. E. 1993, Sol Phys, 144, 217
\bibitem[Dosch et al.(2013)]{Dosch13} Dosch, A., Adhikari, L., \& Zank, G.~P.\ 2013, AIP Conf. Proc., 1539, 155, doi: 10.1063/1.4811011
\bibitem[Fontenla et al.(1999)]{Fontenla1999} Fontenla, J. M., White, O. R., Fox, P. A., Avrett, E. H., \& Kurucz, R. L. 1999, ApJ, 518, 480
\bibitem[Fontenla et al.(2006)]{Fontenla2006} Fontenla, J. M., Avrett, E. H., Thuillier, G., \& Harder, J. 2006,
ApJ, 639, 441
\bibitem[Hazeltine(1983)]{Hazeltine1983} Hazeltine, R. D. 1983, Phys. Fluids, 26, 3242
\bibitem[Kadomtsev \& Pogutse(1974)]{Kadomtsev1974} Kadomtsev, B. B., \& Pogutse, O. P. 1974, Sov. Phys.-JETP, 38, 283
\bibitem[Kryukov et al.(2012)]{Kryukov12} Kryukov, I.~A., Pogorelov, N.~V., Zank, G.~P., et al. \ 2012, AIP Conf. Proc., 1436, 48, doi: 10.1063/1.4723589
\bibitem[Li \& Ding(2009)]{Li2009} Li, Y., \& Ding, M. D. 2009, Res. in Astron. Astrophys., Vol. 9, No. 7, 829
\bibitem[Matthaeus et al.(1996)]{Matthaeus96} Matthaeus, W.~H., Zank, G.~P., \& Oughton, S.\ 1996, JPlPh, 56, 659
\bibitem[Matthaeus et al.(1999)]{Matthaeus99} Matthaeus, W.~H., Zank, G.~P., Oughton, S., et al. \ 1999, ApJL, 523, L93
\bibitem[Montgomery(1982)]{Montgomery1982} Montgomery, D. C. 1982, Phys. Scr., T2/1, 83
\bibitem[Oughton et al.(2001)]{Oughton2001} Oughton, S., Matthaeus, W. H., Dmitruk, P., et al. 2001, ApJ, 551, 565
\bibitem[Parker(1988)]{Parker88} Parker, E.~N.\ 1988, ApJ, 330, 474
\bibitem[Pogorelov et al.(2014)]{Pogorelov14} Pogorelov, N.~V., Borovikov, S.~N., Heerikhuisen, J., et al. \ 2014, in XSEDE'14: Proceedings of the 2014 Annual Conference on Extreme Science and Engineering Discovery Environment, Article No.: 22, 1, (ACM: New York), doi: 10.1145/2616498.2616499
\bibitem[Pontin \& Hornig(2015)]{PH15} Pontin, D.~I., \& Hornig, G.\ 2015, ApJ, 805, 47
\bibitem[Pontin et al.(2017)]{Pontin17} Pontin, D.~I., Janvier, M., Tiwari, S.~K., et al. \ 2017, ApJ, 837, 108
\bibitem[Pontin et al.(2020)]{Pontin20} Pontin, D.~I., Peter, H., \& Chitta, L.~P.\ 2020, A\&A, 639, A21
\bibitem[Powell et al.(1999)]{Powell99} Powell, K.~G., Roe, P.~L., Linde, T.~J., et al. \ 1999, J Comp Phys, 154, 284
\bibitem[Rappazzo \& Parker(2013)]{RP13} Rappazzo, A.~F., \& Parker, E.~N.\ 2013, ApJL, 773, L2
\bibitem[Shebalin et al.(1983)]{Shebalin83} Shebalin, J.~V., Matthaeus, W.~H., \& Montgomery, D.\ 1983, JPlPh, 29, 525
\bibitem[Rosner, Tucker \& Vaiana(1978)]{Rosner1978} Rosner, R., Tucker, W. H., \& Vaiana, G. S. 1978, ApJ, 220, 643
\bibitem[Singh et al.(2018)]{Singh18} Singh, T., Yalim, M.~S., \& Pogorelov, N.~V.\ 2018, ApJ, 864, 18
\bibitem[Title \& Schrijver(1998)]{TS98} Title, A.~M., \& Schrijver, C.~J.\ 1998, in ASP Conf. Ser. 154, Cool Stars, Stellar Systems, and the Sun, ed. R.~A. Sun \& J.~A. Donahue (San Francisco, CA:ASP), 345
\bibitem[van Ballegooijen et al.(2011)]{van Ballegooijen11} van Ballegooijen, A.~A., Asgari-Targhi, M., Cranmer, S.~R., et al. \ 2011, ApJ, 736, 3
\bibitem[Strauss(1976)]{Strauss1976} Strauss, H. R. 1976, Phys. Fluids, 19, 134
\bibitem[van Ballegooijen \& Asgari-Targhi(2016)]{vBTA16} van Ballegooijen, A.~A., \& Asgari-Targhi, M.\ 2016, ApJ, 821, 106
\bibitem[van Ballegooijen \& Asgari-Targhi(2017)]{vBTA17} van Ballegooijen, A.~A., \& Asgari-Targhi, M.\ 2017, ApJ, 835, 10
\bibitem[van Ballegooijen et al.(2017)]{van Ballegooijen17} van Ballegooijen, A.~A., Asgari-Targhi, M., \& Voss, A.\ 2017, ApJ, 849, 46
\bibitem[Warren et al.(2008)]{Warren2008} Warren, H. P., Winebarger, A. R., Mariska, J. T., Doschek, G. A., \&
Hara, H. 2008, ApJ, 677, 1395
\bibitem[Wilmot-Smith(2015)]{Wilmot-Smith15} Wilmot-Smith, A.~L.\ 2015, RSPTA, 373, 20140265
\bibitem[Yalim et al.(2017)]{Yalim17} Yalim, M.~S., Pogorelov, N.~V., \& Liu, Y.\ 2017, J. Phys.: Conf. Series, 837, 012015
\bibitem[Zank \& Matthaeus(1992)]{Zank1992} Zank, G. P., \& Matthaeus, W. H. 1992, J. Plasma Phys., 48, 85
\bibitem[Zank et al.(2012)]{Zank12} Zank, G.~P., Dosch, A., Hunana, P., et al. \ 2012, ApJ, 745, 35
\bibitem[Zank et al.(2017)]{Zank17} Zank, G.~P., Adhikari, L., Hunana, P., et al. \ 2017, ApJ, 835, 147
\bibitem[Zank et al.(2018)]{Zank18} Zank, G.~P., Adhikari, L., Hunana, P., et al. \ 2018, ApJ, 854, 32
\bibitem[Zank et al.(2021)]{Zank21} Zank, G.~P., Zhao, L.-L., Adhikari, L., , et al. \ 2021, Phys. Plasmas 28, 080501; doi: 10.1063/5.0055692
\end{thebibliography}
\end{document}